\documentclass[fleqn,usenatbib]{mnras}

\usepackage{newtxtext,newtxmath}
\usepackage[T1]{fontenc}

\DeclareRobustCommand{\VAN}[3]{#2}
\let\VANthebibliography\thebibliography
\def\thebibliography{\DeclareRobustCommand{\VAN}[3]{##3}\VANthebibliography}

\newcommand{\soutdif}{\bgroup\markoverwith{\textcolor{magenta}{\rule[0.5ex]{2pt}{1pt}}}\ULon}

\usepackage{graphicx}	% Including figure files
\usepackage{amsmath}	% Advanced maths commands
\usepackage{caption}
\usepackage{xcolor}
\usepackage{subcaption}
\usepackage[normalem]{ulem} % For \sout{}
\usepackage{orcidlink}
\title[Impact of Enhanced BH Growth on CGM Clumps]{Feeding or Fleeing: The Impact of Enhanced Early Black Hole Growth on Circumgalactic Clumps at $z > 6$}

\author[F. van Donkelaar et al.] {Floor van Donkelaar$^{\orcidlink{0000-0002-7235-9747}}$,$^{1, 2}$\thanks{E-mail: fv294@cam.ac.uk} Lucas Tortora$^{\orcidlink{0009-0005-8040-8325}}$$^{1, 2}$ and Debora Sijacki$^{\orcidlink{0000-0002-3459-0438}}$$^{1, 2}$ %and %Tiago Costa$^{\orcidlink{0000-0002-6748-2900}}$$^{3}$
\\
$^{1}$Institute of Astronomy, University of Cambridge, Madingley Road, Cambridge CB3 0HA, UK\\
$^{2}$Kavli Institute for Cosmology, Cambridge (KICC), University of Cambridge, Madingley Road, Cambridge CB3 0HA, UK}%\\
\date{MNRAS, to be submitted}

\pubyear{2026}

\begin{document}
\label{firstpage}
\pagerange{\pageref{firstpage}--\pageref{lastpage}}
\maketitle

% Abstract of the paper
\begin{abstract}
The circumgalactic medium (CGM) of massive high-redshift galaxies contains inflowing filamentary gas, recycled material, and feedback-driven outflows. We investigate the formation and evolution of dense CGM clumps in two cosmological zoom-in simulations of the same massive halo at $z > 6$: the fiducial \textsc{Fable} model and the modified \textsc{Fable}-SE model, which includes enhanced early black hole growth through mild super-Eddington accretion. We identify dense gas clumps in the CGM and follow their gas properties, stellar content, and radial evolution across cosmic time. At the resolution probed here, the two simulations show broadly similar distributions of clump gas mass, gas size and associated stellar mass. Instead, the clump populations differ mainly in their abundance, metallicity, and radial evolution. \textsc{Fable}-SE produces substantially more metal-enriched clumps, including an outward-moving population during the strong AGN-driven blow-out phase at $z \lesssim 7$. These systems are associated with disturbed HI-rich gas and relatively high gas densities, linking the increased prevalence of dense, metal-enriched CGM structures in \textsc{Fable}-SE to its stronger AGN-driven blow-out phase. Although inward migration dominates overall, \textsc{Fable}-SE also contains an outward-moving clump population during the blow-out phase, with some clumps moving beyond the virial radius. Our results therefore indicate that the distribution of dense CGM clumps around massive high-redshift galaxies reflects both filamentary accretion and feedback-driven CGM evolution, with strong early AGN activity associated with the outward migration of dense, metal-enriched gas structures containing relatively cool gas.

\end{abstract} %%wordcount = 250


%%%%%%%%%%%%%%%%%%%%%%%%%%%%%%%%%%%%%%%%%%%%%%%%%%
%%%%%%%%%%%%%%%%% BODY OF PAPER %%%%%%%%%%%%%%%%%%

\section{Introduction}\label{sec:introduction}
Galaxies in the early Universe were substantially more gas rich and turbulent than galaxies in the present day \citep[e.g.][]{Saintonge:2013aa, Wisnioski:2015aa, Zanella:2023aa, Semenov:2025aa}. These conditions are also associated with irregular and clumpy morphologies, where massive star-forming clumps are commonly observed within unstable gaseous discs \citep[e.g.][]{Elmegreen:2008aa, Guo:2015aa, Shibuya:2016aa, Harikane:2025aa, Sok:2025aa}. These high gas fractions and turbulent conditions are thought to play an important role in regulating star formation and driving the formation of dense star-forming structures \citep[e.g.][]{Renaud:2021aa, Renaud:2024aa,Donkelaar:2022aa, Ejdetjarn:2022aa}. Recent simulations and observations further suggest that galaxies during cosmic dawn may have experienced highly efficient star formation and rapid stellar mass assembly within compact gas-rich environments \citep[e.g.][]{Dekel:2023aa, Bouwens:2023aa, Harikane:2024aa,vanDonkelaar:2025aa}. Understanding how gas inflows, turbulence and feedback interact in these environments is therefore important for understanding galaxy evolution at high redshift.

At the same time, the circumgalactic medium (CGM) around high-redshift galaxies is also unlikely to be a smooth or passive gas reservoir. Instead, these environments contain a complex mixture of dense inflowing gas, satellites, stripped material, feedback-driven outflows and recycled gas interacting within the same halo. Rapid gas accretion from the cosmic web will continuously feed young galaxies \citep[e.g.][]{keres:2005aa, dekel:2006aa, dekel:2009aa}, while mergers and interactions between satellites further redistribute gas into the surrounding CGM \citep[e.g.][]{stewart:2011aa, angles:2017aa, ginolfi:2020aa}. At the same time, the mentioned high gas fractions and intense star formation rates of these early galaxies drive strong galactic winds capable of ejecting dense and metal-enriched material far beyond the central galaxy 
\citep[e.g.][]{bischetti:2019aa, marshall:2023aa, ginolfi:2020aa}. These different gas components thus coexist within highly dynamic and rapidly evolving environments. Therefore, one may expect that these environments naturally produce compact dense structures embedded within the CGM, including transient star-forming overdensities, stripped clumps, or feedback-compressed gas clouds.

Feedback from rapidly growing black holes (BHs) can drive powerful outflows \citep[for reviews, see][]{Fabian:2012aa, king:2015aa}, which can reshape the high-redshift CGM \citep[e.g.][]{Costa:2015aa, Bennett_et_al_2024}. Simulations of massive quasar hosts at $z > 6$ show that energetic AGN-driven outflows can expel large amounts of metal-rich gas and dust into the halo, enriching and polluting the CGM out to several virial radii \citep[e.g.][]{Sijacki_et_al_2009, Costa_et_al_2014, Costa:2015aa, costa:2022aa, bieri:2017aa, barai:2018aa, zhu:2022aa, ward:2024aa}. These outflows can reach velocities of hundreds to more than $1000~{\rm km~s^{-1}}$ and strongly perturb the surrounding gas halo \citep[e.g.][]{Costa:2015aa, bischetti:2019aa}, reducing central gas densities while simultaneously redistributing metal-rich dense gas throughout the CGM \citep[e.g.][]{costa:2018aa, costa:2020aa}. Rather than simply destroying cold material, recent simulations suggest that AGN-driven feedback can also compress gas within the halo, potentially triggering star formation in outflowing material or long-lived dense clumps \citep[e.g.][]{ishibashi:2012aa, zubovas:2014aa, Costa:2015aa}. Such feedback cycles may therefore strongly influence the formation, disruption, and survival of dense circumgalactic structures \citep[e.g.][]{Costa:2015aa, Nelson:2019aa,
ward:2024aa, Almeida+_2026}.

Observations with the Atacama Large Millimeter/submillimeter Array \citep[ALMA; ][]{Wootten_Thompson_2009} and James Webb Space Telescope \citep[\textit{JWST};][]{Gardner:2006aa} have been beginning to reveal similarly structured environments around massive high-redshift galaxies and quasars \citep[e.g.][]{ding:2023aa, galbiati:2023aa, tripodi:2023aa, bischetti:2025aa}. Extended Ly$\alpha$ nebulae, metal-enriched gas and large-scale outflows indicate that baryons are already being efficiently cycled between galaxies and their surrounding haloes within the first billion years of cosmic time \citep[e.g.][]{ginolfi:2020aa, bischetti:2019aa, marshall:2023aa}. However, observations provide only instantaneous snapshots of these systems. Simulations are therefore important for connecting these different gas phases and for understanding whether dense CGM structures are transient features, disrupted satellites, recycled outflow material, or long-lived clumps that may survive and evolve within the halo environment \citep[e.g.][]{vanDonkelaar:2023aa, Donkelaar:2026aa, ward:2024aa}.

Furthermore, amongst \textit{JWST} most surprising revelations is the discovery of remarkably dense and compact star clusters at redshifts $z \gtrsim 6$--10, many detected through gravitational lensing \citep[e.g.][]{Vanzell:2023aa, Adamo:2024aa, Mowla:2024aa, bradac:2025aa}. These systems exhibit extremely compact sizes and high stellar surface densities, with some showing structural similarities to present-day globular clusters (GCs), while others resemble massive star-forming clumps observed at lower redshift \citep[e.g.][]{Mowla:2022aa, Mowla:2024aa, Adamo:2023aa, Messa:2025aa}. Importantly, these observations suggest that clustered star formation may have represented a fundamental mode of stellar assembly in the early Universe, particularly in the dense and gas-rich environments present during cosmic dawn \citep[e.g.][]{Tacchella:2020aa, Garcia:2023aa,Belokurov:2023aa, Finkelstein:2024aa}.

Similarly, recent numerical simulations indicate that compact stellar systems can form efficiently under a variety of high-redshift conditions. In addition to fragmentation within galactic discs \citep[e.g.][]{dekel:2009aa, Mandelker:2017aa, Renaud:2021aa, Donkelaar:2022aa}, accreting filaments, merger-driven gas flows, and feedback-compressed environments may also trigger the formation of dense stellar systems \citep[e.g.][]{Mandelker:2018aa, vanDonkelaar:2023aa, Mayer:2025aa, Ong:2025aa}. These simulations produce compact clusters with masses and surface densities comparable to those observed with \textit{JWST}, suggesting that multiple interconnected pathways may contribute to clustered star formation during cosmic dawn.

An emerging possibility is that some of these compact systems may form and evolve within the CGM itself, rather than solely inside galactic discs. 
In \citet{Donkelaar:2026aa}, we introduced the term \emph{Cosmic Wallflowers} for compact stellar systems forming outside the main galactic disc, including within the surrounding CGM. Both simulations and observations now hint that dense stellar systems can appear offset from the main stellar body of high-redshift galaxies \citep[e.g.][]{Mowla:2022aa, Claeyssens:2023aa, Giunchi:2025aa, Whitaker:2025aa, vanDonkelaar:2023aa, Donkelaar:2026aa}. Such systems may survive for extended periods in the circumgalactic environment while remaining dynamically disconnected from the central galaxy, forming long-lived offset stellar systems within the halo. Understanding whether these dense CGM clumps fall inward and contribute to galaxy growth, or instead survive and migrate through the halo environment, is therefore important for linking early clustered star formation, feedback, and the baryon cycle around young galaxies.

The survival and evolution of these dense circumgalactic structures is likely sensitive to the strength and modality of BH growth and feedback. In particular, phases of rapid BH growth through super-Eddington accretion can strongly perturb the baryon cycle around young galaxies, driving powerful outflows, redistributing metal-rich gas, and altering the balance between inflowing and outflowing material within the halo \citep[e.g.][]{Volonteri:2010aa, Inayoshi:2020aa, Bennett_et_al_2024, Tortora_et_al_2026}. At the same time, such feedback may also enhance fragmentation and dense structure formation within the CGM itself through compressive turbulence, recycled gas flows, and outflow interactions \citep[e.g.][]{zubovas:2014aa, Costa:2015aa, ishibashi:2012aa}. Whether these environments predominantly destroy dense circumgalactic structures or instead help sustain long-lived offset clusters remains unclear.

In this work, we investigate the properties, evolution and survival of dense CGM gas clumps in two cosmological zoom-in simulations of the same massive halo at $z > 6$: the fiducial \textsc{Fable} model and the modified \textsc{Fable}-SE model, which includes enhanced BH growth through super-Eddington accretion. We identify dense gas clumps directly within the circumgalactic environment and follow their evolution across cosmic time, focusing on their gas properties, stellar content, radial migration and survival. By comparing the two simulations, we explore how different BH growth and feedback modes alter the abundance, properties and fate of dense gas clumps in the CGM during the first billion years of cosmic time.  In Section~\ref{sec:method}, we describe the simulations and clump identification procedure, including the fiducial \textsc{Fable} model (Section~\ref{sec:fable}), the modified \textsc{Fable}-SE model (Section~\ref{sec:fable-se}), and the clump selection methodology (Section~\ref{sec:identifiy_clump}). In Section~\ref{sec:results}, we present the properties and evolution of the identified CGM clumps, focusing on their gas properties (Section~\ref{sec:gas_properties}), stellar content (Section~\ref{sec:stellar}), and radial migration (Section~\ref{sec:radial_migration}). Finally, we discuss the implications of our findings and summarize our conclusions in Section~\ref{sec:disc}.

\section{Methods}\label{sec:method}
\subsection{Simulation codes \& Initial conditions}
% This work uses zoom-in simulations that are based on the \textsc{fable} suite of cosmological simulations of (clusters of) galaxies \citep{Henden:2018aa, Bigwood_et_al_2025}. 
This work uses cosmological, zoom-in simulations run with the \textsc{arepo} gravity and hydrodynamics code \citep{Springel_2010, Pakmor_et_al_2016, Weinberger_et_al_2020}. The code discretizes and solves the equations of hydrodynamics on a moving mesh constructed from the Voronoi tessellation of a discrete set of mesh-generating points, using a quasi-Lagrangian, finite volume approach. \textsc{arepo} keeps cells within a factor of 2 of a target mass $m_{\rm target}$, (de-)refining as needed \citep[][]{Springel_2010}. Gravitational interactions are computed via a tree-particle method. 

We analyse `zoom-in' simulations targeting the most massive halo in the Millennium cosmological volume at $z=6$ \citep[][]{Springel_et_al_2005b, Sijacki_et_al_2009}. 
We generate high-resolution initial conditions at $z=127$ by selecting the Lagrangian patch evolving to the target halo at $z \, = \, 6$, and populating it with high-resolution cells with target mass $m_{\rm target}=1.99\times10^6~{\rm M_{\sun}}$. The resolution outside this region is then progressively deteriorated, increasing the cell target mass at larger distances, while capturing the large-scale cosmological tidal field. Further details of the re-simulation technique can be found in \citet{Sijacki_et_al_2009}. 

At $z=6$, the target halo has a  virial\footnote{In this work, we use the spherical overdensity criterion to define virial quantities, i.e. $M_{\rm vir}$ is the mass contained within $R_{\rm vir}$, which is the radius within which the average density is 200 times the critical density of the universe, at a given redshift.} mass and radius of $M_{\rm vir}\approx 7 \times10^{12}~{\rm M_\odot}$ and $R_{\rm vir}\approx 85~{\rm kpc}$, respectively. The mass resolution for gas cells and dark matter particles is $1.99\times10^6~{\rm M_\odot}$ (i.e., $m_{\rm target}$) and $9.07\times10^6~{\rm M_\odot}$, respectively. The softening lengths for dark matter and stars are $\epsilon_{\rm soft}=2.5~{\rm comoving \ kpc}$. For gas cells, the gravitational softening is adaptive and reaches $\sim0.2~{\rm kpc}$ within 2~kpc from the halo centre at $z=6$.

The same cosmological parameters from the parent Millennium simulation are used, which are consistent with \textit{WMAP} constraints \citep{Spergel_et_al_2007}: $\Omega_\Lambda=0.75, \, \Omega_{\rm m}=0.25, \, \Omega_{\rm b}=0.045, \, \sigma_8 = 0.9$ and $h=0.73$. We refer to \citet{Bennett_et_al_2024, Tortora_et_al_2026} for further details on the setup of the `zoom-in' simulations.

\subsubsection{The \textsc{fable} model} \label{sec:fable}
As in any other cosmological simulation, we employ sub-grid models to capture the physics of galaxy formation, including gas cooling, star formation, stellar feedback, and the seeding, accretion and feedback from BHs. In this work, we use variations of the \textsc{fable} model \citep[][]{Henden:2018aa, Bigwood_et_al_2025}, which is a modified version of the Illustris model \citep[][]{Vogelsberger_et_al_2013, Vogelsberger_et_al_2014, Genel_et_al_2014, Torrey_et_al_2014, Sijacki:2015aa}; we provide a brief overview of the model here, and refer the interested reader to the aforementioned works for further details.

% Star formation
The interstellar medium (ISM) of the simulations is unresolved and modelled using an effective equation of state (eEoS) in a sub-grid fashion, following \citet{Springel_Hernquist_2003} and the modifications introduced in \citet{Vogelsberger_et_al_2013}. Star formation occurs stochastically in gas with density above a threshold $n_{\rm sfr}$ with a characteristic time scale $t_{\rm sfr}$. The adopted parameters for the simulations are $n_{\rm sfr}=0.13~{\rm cm^{-3}}$ and $t_{\rm sfr}=2.2~{\rm Gyr}$. Stellar particles inherit the metallicity of their parent gas cells. Each stellar particle represents a single-age stellar population with the \citet{Chabrier_2003} initial mass function. The thermal regulation of the unresolved, star-forming ISM by supernovae is incorporated implicitly through the eEoS model \citep{Springel_Hernquist_2003, Vogelsberger_et_al_2013}. Explicit stellar feedback is implemented through the stochastic launching of galactic winds from star-forming gas. Wind particles are temporarily decoupled from hydrodynamic interactions, subsequently depositing their mass, metals, momentum and thermal energy into the surrounding gas. Note that one third of this wind energy is injected as thermal rather than kinetic energy in the simulations, to slow down the cooling of ejected gas, thus maintaining it outside of the ISM for longer periods of time and increasing the effectiveness of stellar feedback \citep[e.g.,][]{Marinacci_et_al_2014, Henden:2018aa}.

% Black hole accretion
Black holes are treated as collisionless sink particles that may grow either through accretion or mergers with other BHs. A BH particle with mass $M_{\rm BH, \, seed}=10^{5}~{h^{-1} \, \rm M_\odot}$ is `seeded' at the position of the gravitational potential minimum of every dark matter halo with mass $M_{\rm vir, \, seed}>5\times10^{10}~{h^{-1} \, \rm M_\odot}$. Accretion of gas onto BHs is parametrised by the Bondi-Hoyle-Lyttleton model \citep{Hoyle_Lyttleton_1939, Bondi_Hoyle_1944, Bondi_1952}, where the accretion rate is given by:
\begin{equation}\label{eq:BHL}
    \dot{M}_{\rm BH}=\alpha \frac{4\pi \, G^2 \, M_{\rm BH}^2 \, \rho}{c_{s}^3}\,,
\end{equation}
where $\dot{M}_{\rm BH}$ is the BH accretion rate, $G$ is the gravitational constant, $M_{\rm BH}$ is the BH mass, $\rho$ and $c_s$ are the density and sound speed of the surrounding gas, and $\alpha$ is a numerical boost factor (in \textsc{fable}, we use $\alpha=100$). This is introduced to numerically boost the accretion of gas onto BHs because the cold, dense phase is not explicitly resolved, which could feed the BH at much higher rates than numerically captured \citep[see e.g.][for discussions]{Springel_et_al_2005a}.

% Eddington limit
The gas feeding the BH also ignites an AGN \citep[e.g.,][]{Rees_1984}, whose luminosity is simply given by $L_{\rm bol}=\epsilon_{\rm r} \, \dot{M}_{\rm BH} \, c^2$, with $\epsilon_{\rm r}=0.1$ representing the radiative efficiency of BHs in \textsc{fable}. The accretion rate is assumed to be capped at the Eddington limit as a result of this radiative output, where:
\begin{equation}\label{eq:edd-limit}
    \dot{M}_{\rm Edd}=\frac{4\pi \, G\, M_{\rm BH} \, m_{\rm p}}{\epsilon_{\rm r} \, \sigma_{\rm T}\, c}\,,
\end{equation}
with $m_{\rm p}$ the proton mass, $c$ the speed of light and $\sigma_{\rm T}$ the Thomson scattering cross-section (assuming ionized hydrogen here). In \textsc{fable}, the Eddington rate is such that $f_{\rm Edd}:=\dot{M}_{\rm BH} \, / \, \dot{M}_{\rm Edd} \leq 1$. A pressure criterion further modulates the accretion rate during epochs when large BHs are embedded in low-density gas, to avoid the unphysical inflation of hot and diffuse bubbles due to feedback \citep[][]{Vogelsberger_et_al_2013}.

% Feedback from AGN
Feedback from AGN is simulated using a dual-mode feedback model depending on the accretion rate. At high accretion rates $f_{\rm Edd}>0.01$, the (radiatively efficient) luminous quasar mode operates and 10 per cent of the bolometric luminosity powering the AGN is isotropically injected as thermal energy in the BH neighbourhood \citep[][]{Springel_et_al_2005a}. \textsc{fable} imposes a duty cycle of 25~Myr, during which the feedback energy is accumulated and stored before being released in a single event. This is done to mitigate the likely artificial overcooling in dense gas in the vicinity of the BH \citep[see][]{Booth_Schaye_2009}. At lower accretion rates $f_{\rm Edd}\leq0.01$, the (radiatively inefficient) radio mode operates, which injects hot bubbles into the gas as if inflated by jets. Injection occurs in bursts after the BH mass changes by 1 per cent \citep{Sijacki_et_al_2007}. In the Illustris and the original \textsc{fable} simulations, all remaining energy that is not coupled to the gas through feedback (in both the quasar and radio modes) goes into radiative feedback \citep[but does not self-consistently track AGN radiation through radiative transfer, see][for details]{Vogelsberger_et_al_2013, Henden:2018aa}. This feature is turned off in our simulations and would require on-the-fly radiative transfer to be properly investigated. We demonstrate the impact of including quasar events with radiative transfer performed in post-processing in \citet{Tortora_et_al_2026}.

% Dynamics of BH particles
Dynamical friction is not accurately resolved in most cosmological simulations due to limited mass and spatial resolution. In \textsc{fable}, BHs are thus kept at the centre of their host halo by artificially migrating them to the minimum gravitational potential of particles within their smoothing lengths at every gravitational time step \citep[see][]{Springel_et_al_2005a}. Additionally, two BHs instantaneously merge when they are within each other's smoothing lengths. Combined, these effects produce overly efficient and abundant mergers \citep[see discussions in][]{Buttigieg_et_al_2025}.

\subsubsection{Modifications: the \textsc{fable}-SE model} \label{sec:fable-se}

% General
The \textsc{fable}-SE model introduces changes to the fiducial \textsc{fable} model to promote early BH growth and delay the onset of efficient black hole self-regulation through feedback \citep[][]{Bennett_et_al_2024}. Other model parameters, related to gas cooling, star formation and stellar feedback, are kept identical. The modifications pertain only to BH growth and feedback models. These are summarised below.

% Seeding change
We seed BHs earlier by reducing the halo mass threshold for BH seeding in \textsc{fable}-SE to $M_{\rm vir, \, seed}>10^{9}~{h^{-1} \, \rm M_\odot}$. This increases the redshift at which the first BH is seeded, effectively gaining around $\sim100$~Myr for it to grow through gas accretion. This change also significantly increases the total number of BHs in the simulations. While this is noted to have non-trivial consequences on the early mass assembly and growth history of BHs, we note that this does not significantly affect the final BH mass, which is mostly gained through accretion \citep[see discussions in][]{Sijacki_et_al_2009, Bhowmick_et_al_2022, Bennett_et_al_2024}.

% Accretion & feedback changes
In \textsc{fable}-SE, we increase the cap on the maximal accretion rate of BHs and allow mildly super-Eddington accretion of gas up to $f_{\rm Edd}=2$.
A number of changes are also made to the feedback model and parameters in \textsc{fable}-SE. The radio mode feedback at low accretion rates is switched off entirely, which does not significantly alter BH growth given that the quasar mode is dominant at high redshifts \citep[][]{Bennett_et_al_2024}. The duty cycle in the quasar mode feedback is turned off, and we inject feedback energy continuously in \textsc{fable}-SE. Finally, the coupling strength of feedback in the quasar mode is reduced to 5 per cent, down from the 10 per cent used in \textsc{fable}.

% Mini-conclusion
These changes combine to produce an ultramassive BH at $z=6$ in \textsc{fable}-SE ($M_{\rm BH, \, \textsc{fable}-SE}=1.3\times10^{10}~{\rm M_\odot}$) that is one order of magnitude higher than in \textsc{fable} ($M_{\rm BH, \, \textsc{fable}}=1.4\times10^{9}~{\rm M_\odot}$). While the modifications promote BH growth in part by reducing the AGN feedback efficiency at very high redshifts, the total injected energy at $z = 6$ depends on the BH mass and is thus significantly higher in \textsc{fable}-SE \citep[][]{Tortora_et_al_2026}. As a result, the BH in \textsc{fable}-SE drives much more powerful outflows that completely alter the properties and structure of the host halo \citep[][]{Bennett_et_al_2024, Tortora_et_al_2026}, which will be the focus of the present study. We note that \textsc{Fable}-SE does not introduce a separate feedback mode above the Eddington limit. The stronger blow-out phase in \textsc{Fable}-SE is therefore driven by earlier and more rapid BH growth, rather than by a distinct super-Eddington feedback model.

\subsection{Identifying Clumps} \label{sec:identifiy_clump}
We identify clumps directly from the gas distribution at each snapshot. In this study our aim is to study overdense gas structures in the CGM, rather than to identify individual stellar clusters, or individual self-gravitating structures. At the resolution of these simulations, such overdensities may represent larger gas complexes containing unresolved substructures, and are not necessarily gravitationally bound or physically strictly distinct systems. We do not explicitly exclude gas associated with satellite galaxies, meaning that some identified clumps may be within satellites. We examine the possible contribution from satellite galaxies in Section~\ref{sec:clumpsat}, where we find that the additional outward-moving population in \textsc{Fable}-SE, particularly for clumps born at $z_{\rm birth}<7$, is predominantly not associated with resolved satellites. Throughout this work, the term CGM clump therefore refers to our gas-selected population, rather than necessarily to physically distinct structures in the CGM.

\begin{figure*}
    \centering
        \includegraphics[ trim={0cm 0cm 0cm 0cm}, clip, width=0.99\textwidth, keepaspectratio]{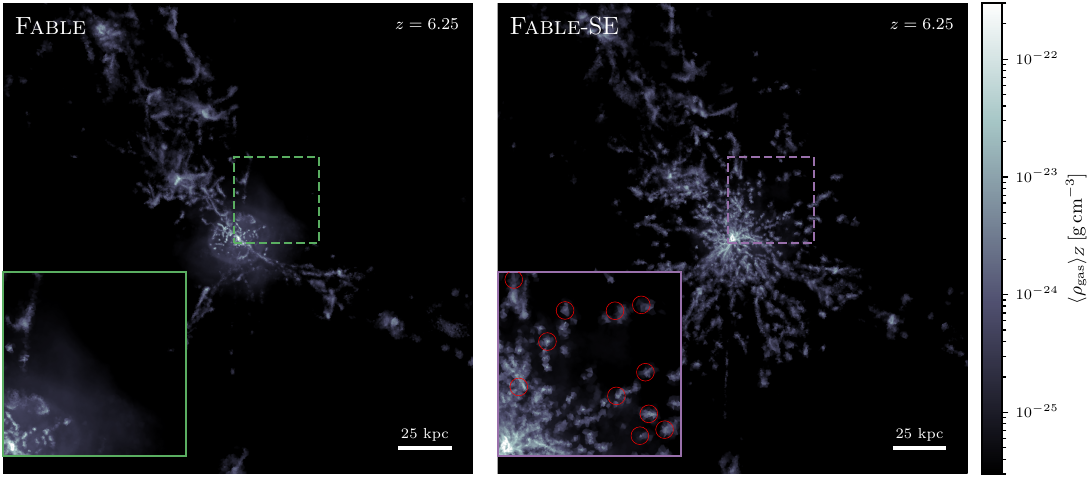}
    \caption{Metallicity-weighted gas density projections of FABLE (left) and FABLE-SE (right) at $z=6.25$. Each panel covers $220 \times 220$ physical kpc, with a line-of-sight depth of $220$ physical kpc. The colour scale shows $\langle \rho_{\rm gas} \rangle_Z = \int \rho_{\rm gas}^2 Z\,dl / \int \rho_{\rm gas} Z\,dl$ \citep[see also][]{Tortora_et_al_2026}. The dashed squares indicate the $40 \times 40$ physical kpc regions shown in the enlarged insets. Red circles mark gas clumps containing stars that are tracked for at least five subsequent snapshots.}
    \label{fig:main}
\end{figure*} 

We use the friends-of-friends (FoF) and subhalo catalogues to define the main system. The FoF catalogue groups particles into haloes, while the subhalo catalogue identifies gravitationally bound structures, i.e. galaxies within those haloes. At each snapshot, we take the most massive FoF group as the main halo and define the central galaxy as the most massive subhalo within this group. All galactocentric distances are measured relative to the position of minimum gravitational potential of the central subhalo, as given by the \texttt{SubhaloPos} catalogue field.

The clump search is restricted to gas cells at galactocentric distances of $25~{\rm kpc} < r < 105~{\rm kpc}$. We adopt an inner cut of $25$~kpc, well outside the central galaxy's stellar half-mass radius of $\lesssim 5$~kpc, to exclude the central galaxy and its immediate surroundings and focus on the CGM. We first select gas above an overdensity threshold relative to the mean baryon density at that redshift,
\begin{equation}
    \rho_{\rm gas}/\bar{\rho}_{\rm b} \geq 200 \,.
\end{equation}

We adopt this threshold to select gas substantially denser than the cosmological background while retaining dense CGM structures below the simulation's star-formation threshold. At $z=6$, it corresponds to $n_{\rm H}\simeq1.4\times10^{-2}~{\rm cm}^{-3}$, approximately an order of magnitude below the star-formation threshold. We test a higher threshold of 500 in Appendix~\ref{app:clump_finder}.

A FoF algorithm is then applied to these dense gas cells. We adopt a linking length of $1~{\rm kpc}$, grouping neighbouring overdense cells into larger gas structures without imposing a fixed shape. Our aim is to identify overdense regions in the CGM that may contain unresolved substructures, potentially including sites of compact stellar cluster formation, rather than individual bound clouds. Using a shorter linking length identifies smaller and denser structures, but at $0.25~{\rm kpc}$ substantially fewer clumps satisfy our minimum cell-count criterion. We therefore adopt $1~{\rm kpc}$ to retain larger gas complexes, while exploring the dependence on the linking length and other clump-finding parameters in Appendix~\ref{app:clump_finder}. Gas outside the initial dense-cell selection is not used to define the FoF groups, but is included afterwards when measuring quantities within the gas half-mass radius around the clump centre.

We define the centre of each clump as the centre of mass of its member gas cells,
\begin{equation}
    \mathbf{x}_{\rm clump} =
    \frac{\sum_i m_i \mathbf{x}_i}{\sum_i m_i},
\end{equation}
where $m_i$ and $\mathbf{x}_i$ are the mass and position of gas cell $i$. We use this centre to measure the gas half-mass radius and the clump's position relative to the central galaxy; it is not the position of the most bound or densest cell.

As our default choice we retain clumps containing at least 16 gas cells (to avoid poorly resolved structures), with a total gas mass above $10^6~{\rm M_\odot}$, and with gas half-mass radii in the range $0.1~{\rm kpc} < R_{\rm half,gas} < 15~{\rm kpc}$\footnote{We apply the same half-mass radius limits and clump selection definitions to both simulations to ensure a consistent comparison.}. The sensitivity to linking length and overdensity threshold is illustrated in Appendix~\ref{app:clump_finder}, while the dependence on clump identification parameters is discussed further in Section~\ref{sec:res}.

These cuts restrict the catalogue to relatively compact, massive gas structures, while removing very extended overdensities that are unlikely to correspond to individual clumps. Unless stated otherwise, property distributions count detections across snapshots, so a tracked clump may contribute more than once.

We also measure the internal gas motions of the identified clumps. For this, we use an aperture of $R_{\rm ap}=\max(R_{\rm half,gas},0.2\,{\rm kpc})$ and calculate the mass-weighted one-dimensional gas velocity dispersion, $\sigma_{\rm 1D}$. We estimate the virial parameter as

\begin{equation}
    \alpha_{\rm vir} = \frac{5\sigma_{\rm 1D}^2 R_{\rm ap}}{G M_{\rm gas}(<R_{\rm ap})}\,,
\end{equation}

where $M_{\rm gas}(<R_{\rm ap})$ is the gas mass enclosed within the aperture. To characterise local converging and expanding gas motions, we additionally define

\begin{equation}
    C = \frac{\sum_{v_{r,i}<0} m_i |v_{r,i}|}
    {\sum_{v_{r,i}>0} m_i v_{r,i}},
\end{equation}

where $v_{r,i}$ is the radial velocity of each gas cell relative to the clump centre and its bulk velocity. Values of $C>1$ indicate that the mass-weighted converging motions dominate over expanding motions. Both quantities provide approximate diagnostics of gas dynamics within clumps, rather than direct measurements of gravitational binding or collapse. 

We further estimate the velocity divergence, $\nabla\cdot\mathbf{v}$, from a mass-weighted linear least-squares fit to the three-dimensional gas velocity field within $R_{\rm ap}$, relative to the clump's bulk velocity. We take the trace of the fitted velocity-gradient tensor, requiring at least eight gas cells. Negative values indicate locally converging gas motions. We also compare the mass- and volume-weighted effective gas pressure, $P=(\gamma-1)\rho u$, within $R_{\rm ap}$ to that of the surrounding CGM\footnote{We use spherical shells at $1.5$--$4R_{\rm ap}$ and $1$--$1.5R_{\rm ap}$, requiring at least eight gas cells in each region. For star-forming gas, this represents the effective pressure prescribed by the eEoS.}.

\begin{figure*}
    \centering
        \includegraphics[ trim={0cm 0cm 0cm 0cm}, clip, width=0.98\textwidth, keepaspectratio]{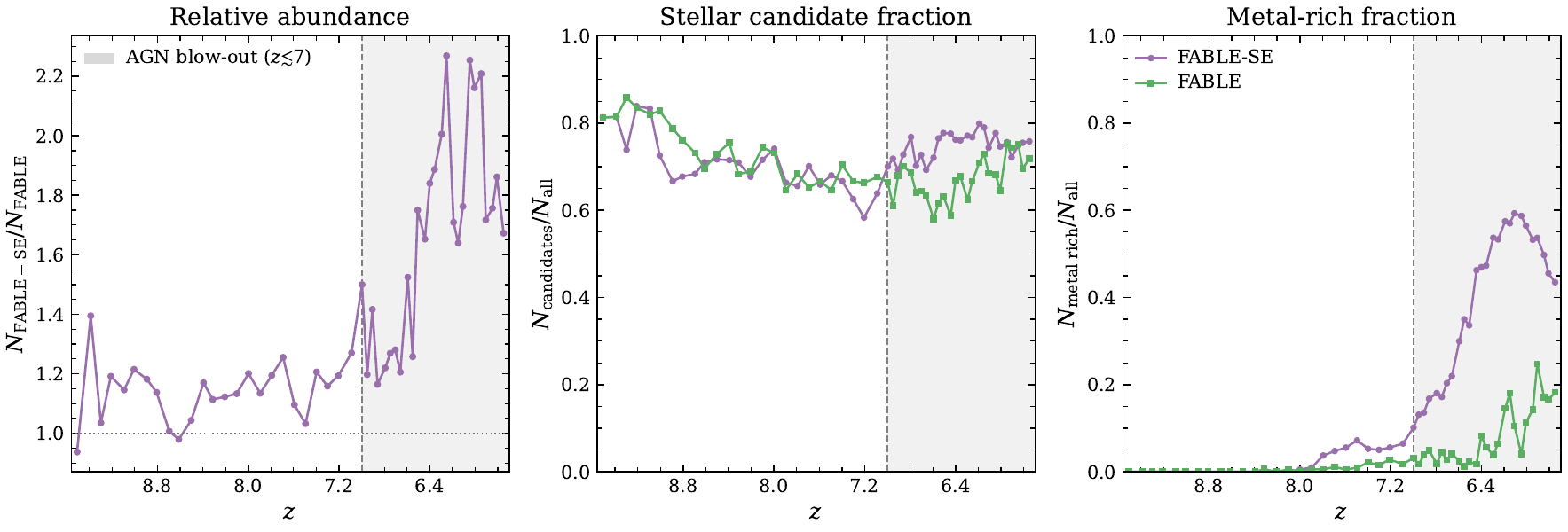}
    \caption{Redshift evolution of the relative abundance of CGM clumps in \textsc{Fable}-SE compared to \textsc{Fable} (left), the fraction of clumps classified as stellar candidates (middle), and the fraction of metal-rich clumps (right). We define metal-rich clumps as those with $\log_{10}(Z_{\rm gas}/Z_\odot) \geq -0.5$. The horizontal dotted line in the left panel marks equal clump abundances. The grey shaded region indicates the strong AGN-driven blow-out phase at $z \lesssim 7$ in \textsc{Fable}-SE.}
    \label{fig:abundance}
\end{figure*}

\subsubsection{Stellar candidates}\label{sec:stellar_candidates}
As described above, the clump catalogue is selected from the gas alone. Stellar properties are therefore attached only after the gas clumps have been identified. For each clump, we select star particles within $R_{\star,\mathrm{ap}}=\max(R_{\mathrm{max,gas}},2R_{\mathrm{half,gas}},0.5~{\rm kpc})$ of the gas-defined centre, where $R_{\mathrm{max,gas}}$ is the maximum distance of a member gas cell from that centre. This keeps the initial selection independent of the stellar content, while allowing us to distinguish purely gaseous clumps from systems that host stars. However, we caution that the presence of stellar particles does not necessarily imply that the gas overdensity corresponds to a gravitationally bound stellar cluster.

For the analysis in this work, we define a subset of \emph{stellar candidates}. These are clumps that either already contain stars or, if starless, contain gas close to the eEOS. We include the latter to identify dense, gas-dominated systems that have not yet formed stars. We define these starless candidates as objects with no stellar component and a volume-averaged hydrogen number density of $n_{\rm H} \geq 0.1~{\rm cm}^{-3}$, measured within $R_{\rm ap}$. This threshold is close to, but slightly below, the gas-cell star-formation threshold of $n_{\rm H}=0.13~{\rm cm}^{-3}$ adopted in the simulations. Because this density is averaged over gas cells within $R_{\rm ap}$, it does not require every cell to exceed the star-formation threshold. We use this criterion as a selection proxy for dense starless gas, not as evidence that the clump as a whole is forming stars or will necessarily do so. The volume-averaged density can miss localised dense gas, and its value depends on which cells are grouped into a clump. Our clump-finding tests show that the stellar-candidate fraction also depends on the adopted linking length. In the four snapshots examined, all clumps retained with linking lengths of $0.25$ and $0.5$~kpc satisfy our stellar-candidate criterion, while the fiducial $1$~kpc selection also includes gas clumps that do not satisfy this criterion. The stellar-candidate population is therefore sensitive to the adopted clump identification parameters. For the virial-radius analysis, we first construct the full gas-clump tracks and then select those satisfying the stellar-candidate criterion in at least one snapshot. The first and last positions are measured from the complete gas track, regardless of whether the candidate criterion is satisfied at these endpoints.

For clumps that already contain stars, we use the measured stellar quantities directly. For starless systems selected through the gas proxy, we simply estimate the stellar mass as
\begin{equation}
    M_{\star,{\rm est}} = \epsilon_\star M_{\rm gas}\,,
\end{equation}
where $\epsilon_\star=0.05$ is an illustrative fiducial conversion efficiency. We explore values of 1, 5 and 10 per cent in Appendix~\ref{sec:SFproxy}. This assumption affects only the estimated stellar properties of starless candidates and does not enter the stellar-candidate selection or gas-clump tracking. For starless candidates, we estimate the stellar surface density using the assumed stellar mass and setting the stellar half-mass radius equal to the gas half-mass radius. We define the surface-density estimators as $\Sigma_{\rm gas}=M_{\rm gas}/(\pi R_{\rm half,gas}^{2})$ and $\Sigma_\star=M_\star/(\pi R_{\rm half,\star}^{2})$.

Figure~\ref{fig:main} shows metallicity-weighted gas density projections of the simulated haloes in \textsc{Fable} and \textsc{Fable}-SE at $z=6.25$, together with zoomed-in regions. The red circles mark the centres of stellar candidates selected from the three-dimensional gas clump catalogue. At this redshift, \textsc{Fable}-SE contains visibly more such clumps than \textsc{Fable}, distributed throughout the CGM and extending to approximately the virial radius.

Note that the catalogue identifies coherent gas structures at the resolution of the simulations, but does not resolve in detail their internal properties or compact stellar substructure. The reported radii and surface densities should therefore be interpreted as large-scale clump properties rather than resolved cluster-scale measurements.

\subsubsection{Tracking between snapshots} %mention tracker particles for future
\label{sec:tracking}
We track clumps between consecutive snapshots by comparing the identifiers of their constituent \textsc{Arepo} mesh cells. A match is accepted when at least 5 per cent of the identifiers in the current clump or 2 per cent of those in the previous clump are shared. We apply the same criteria to both simulations. These identifiers belong to moving mesh elements rather than Lagrangian tracer particles. Since gas can flow between mesh cells, and the mesh can undergo refinement and derefinement, our method follows the continuity of identified gas overdensities rather than identical gas parcels. We test the robustness of this approach in Appendix~\ref{app:tracking}.

\begin{figure*}
    \centering
        \includegraphics[ trim={0cm 0cm 0cm 0cm}, clip, width=0.98\textwidth, keepaspectratio]{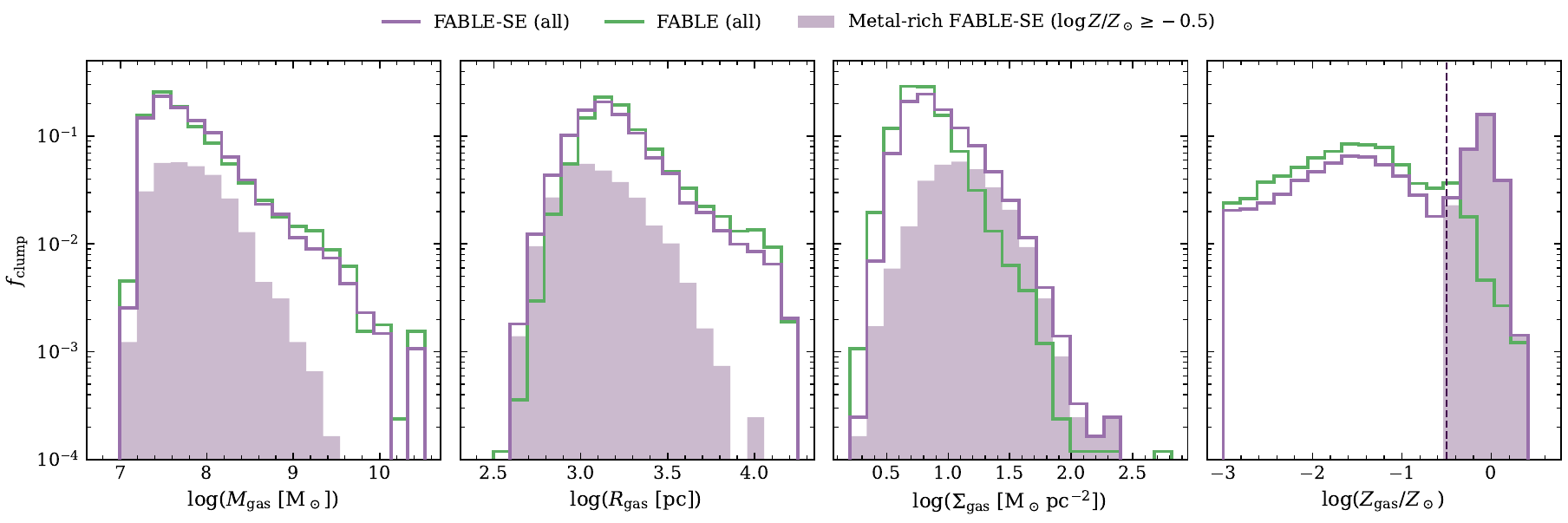}
    \caption{Distribution of gas clump properties in the \textsc{Fable}-SE (purple) and \textsc{Fable} (green) simulations. The panels show the distributions of gas mass, gas half-mass radius, gas surface density ($\Sigma_{\rm gas}$), and gas metallicity. The shaded purple histograms highlight the metal-rich population in \textsc{Fable}-SE, defined by $\log_{10}(Z_{\rm gas}/Z_\odot) \geq -0.5$. The shaded distributions are normalised to the total number of \textsc{Fable}-SE clumps, showing their contribution to the full population.}
    \label{fig:gasprop}
\end{figure*}

For each current clump, we identify the dominant progenitor using the largest number of shared mesh-cell identifiers. When several current clumps select the same progenitor, only the descendant with the highest matching score, $2f_{\rm curr}+f_{\rm prev}$, inherits the original track ID. Here, $f_{\rm curr}$ and $f_{\rm prev}$ are the overlap fractions relative to the current and previous clumps, respectively. The remaining descendants start new tracks and are flagged as split children.

We identify mergers when additional progenitors contribute at least 5 per cent of the identifiers in the current clump\footnote{Discussion on this threshold can be found in Appendix~\ref{app:tracking}}. Clumps without a qualifying match start new tracks. We also visually inspect selected tracks, including mergers and splits, to assess their continuity between snapshots.

\section{Results}\label{sec:results}

\begin{figure}
    \centering
        \includegraphics[ trim={0cm 0cm 0cm 0cm}, clip, width=0.48\textwidth, keepaspectratio]{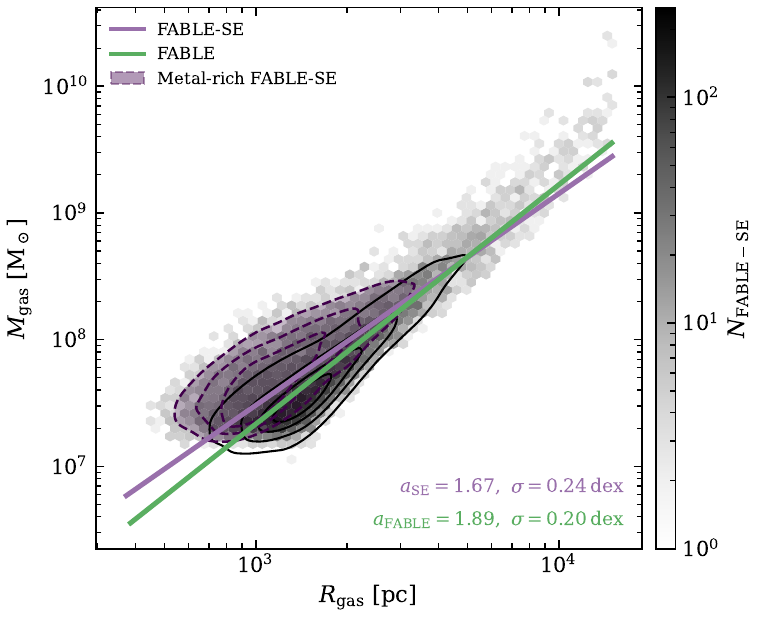}
    \caption{The gas mass, $M_{\rm gas}$, as a function of gas half-mass radius, $R_{\rm gas}$. Grey hexagonal bins show the number density of \textsc{Fable}-SE clumps, with the corresponding colour scale indicated on the right. Black contours show the distribution of \textsc{Fable} clumps, while the purple shaded regions and dashed contours highlight the metal-rich population in \textsc{Fable}-SE, defined by $\log_{10}(Z_{\rm gas}/Z_\odot) \geq -0.5$. The metal-rich contours are normalised to the peak density of that subset. Solid lines indicate the best-fitting power-law relations for the full \textsc{Fable}-SE (purple) and \textsc{Fable} (green) populations. The labels give the fitted slopes, $a$, and the scatter in $\log M_{\rm gas}$ around each relation.}
    \label{fig:larson}
\end{figure}

We identify the dense gas clumps in the CGM of both simulations and separate them into all clumps, and stellar candidates. We first compare the abundance of these populations, before turning to their gas properties, stellar content and radial evolution.

Figure~\ref{fig:abundance} shows the relative abundance of CGM clumps in \textsc{Fable}-SE compared to \textsc{Fable} (left), the fraction of stellar candidates (middle), and the fraction of metal-rich clumps (right). At early times, both simulations contain broadly similar numbers of clumps, with \textsc{Fable}-SE generally having slightly more. At lower redshift, the populations diverge, with the relative abundance increasing strongly in \textsc{Fable}-SE, reaching approximately twice as many clumps around $z \sim 6$.

The difference becomes strongest at $z \lesssim 7$, coinciding with the strong AGN-driven blow-out phase in \textsc{Fable}-SE, marked by the grey shaded region in Figure~\ref{fig:abundance}. During this period, powerful outflows disrupt the central galaxy and CGM, redistributing gas into dense structures and ejecting cold material into the halo \citep[see][]{Tortora_et_al_2026}. The fraction of stellar candidates also becomes generally higher in \textsc{Fable}-SE, although the difference remains relatively small. The clearest difference is in the metal-rich population, which increases strongly in \textsc{Fable}-SE after $z \sim 7$, while remaining much smaller in the fiducial simulation. This suggests that the stronger feedback is particularly associated with an increase in dense, metal-enriched CGM structures.

\subsection{Gas Properties} \label{sec:gas_properties}
Now that we have quantified the abundance of the clumps, we compare their gas properties in the two simulations. Figure~\ref{fig:gasprop} shows the distributions of clump gas mass, gas half-mass radius, gas surface density and gas metallicity for the full sample. The clump gas masses and sizes are broadly similar in \textsc{Fable} and \textsc{Fable}-SE. In both simulations, most clumps have gas masses of $M_{\rm gas}\sim10^{7}$--$10^{8.5}~{\rm M_\odot}$ and gas half-mass radii of order $R_{\rm gas}\sim10^{3}$~pc, with extended tails to both higher masses and larger radii.

The similar gas mass and size distributions in \textsc{Fable} and \textsc{Fable}-SE shows that the additional clumps in \textsc{Fable}-SE do not occupy a completely different range of resolved masses or sizes. The main difference appears in the metallicity distributions, where \textsc{Fable}-SE shows an additional peak close to solar metallicity. Selecting clumps with $\log_{10}(Z_{\rm gas}/Z_\odot) \geq -0.5$ ($Z_{\rm gas} \geq 0.32\,Z_\odot$), we find that this metal-rich population accounts for 43 per cent of clump detections at $z < 7$ in \textsc{Fable}-SE, compared to only 9 per cent in \textsc{Fable}.

The metal-rich clumps in \textsc{Fable}-SE occupy a similar broad mass and size range as the full population, but are generally slightly more massive, smaller, and denser. Restricting the comparison to $z < 7$, their median gas mass is $6.0\times10^7\,{\rm M_\odot}$ compared to $4.5\times10^7\,{\rm M_\odot}$ for the remaining clumps, while their median gas half-mass radius is $1.2$~kpc compared to $1.5$~kpc. Their median gas surface density is approximately twice as high. This additional metal-rich population becomes particularly prominent during the strong AGN-driven blow-out phase, consistent with the redistribution of enriched gas throughout the CGM.

Figure~\ref{fig:larson} shows the corresponding clump gas mass-size relation. Both simulations occupy the same broad region in this plane, confirming that their clumps have similar characteristic masses and radii. We fit a power-law relation between gas mass and gas radius and find slopes of $a = 1.67$ for \textsc{Fable}-SE and $a = 1.89$ for \textsc{Fable}, with scatters of 0.24 and 0.20 dex, respectively. The difference is modest, although \textsc{Fable} shows a slightly steeper mass-size relation. The metal-rich population in \textsc{Fable}-SE occupies a similar broad region of the mass-size plane, although it is more concentrated towards smaller radii. This further suggests that the additional metal-rich clumps do not form a completely separate population in terms of their resolved masses and sizes.

The fitted slopes are relatively close to the scaling $M_{\rm gas}\propto R_{\rm gas}^{2}$, corresponding to approximately constant gas surface density, as commonly discussed in the context of Larson's relations for molecular clouds \citep{larson81, Solomon1987}. However, subsequent studies have shown that molecular cloud surface densities are not necessarily universal and that the inferred mass--size relation can depend on how clouds are defined and selected \citep{heyer_etal09,BallesterosParedes:2012aa}. Our identified structures also differ substantially from molecular clouds in the Milky Way. They are more massive and extended, reside within the turbulent and dynamically evolving CGM of a massive high-redshift galaxy, and are not necessarily gravitationally bound. Moreover, the measured relation may partly reflect our adopted overdensity threshold, linking length and finite numerical resolution. We therefore do not interpret the similarity in slopes as evidence that these systems follow the same underlying physical relations as local molecular clouds.

At the resolution of our simulations, the identified clumps represent extended gas structures rather than individual compact clouds. The gas mass resolution is $1.99\times10^6\,{\rm M_\odot}$; gravitational softening is adaptive for gas, while dark matter and stars have comoving softening lengths of $2.5~{\rm kpc}$ (see Section~\ref{sec:method}). The simulations therefore do not resolve the internal fragmentation or collapse of these dense structures. As a result, the measured clump sizes and inferred surface densities are likely affected by the finite resolution, especially in the CGM. Future work with higher-resolution simulations will be needed to determine whether these structures fragment or contract to smaller radii, potentially revealing more compact substructures within the larger gas complexes identified here.

To summarize, the gas properties show that both simulations produce CGM clumps with broadly similar gas masses and resolved sizes. The main effect of \textsc{Fable}-SE is therefore not a clear shift in the characteristic resolved scale of individual clumps, but an increase in their abundance and the presence of a more metal-rich subset of the population.

\begin{figure}
    \centering
    \includegraphics[width=\columnwidth]{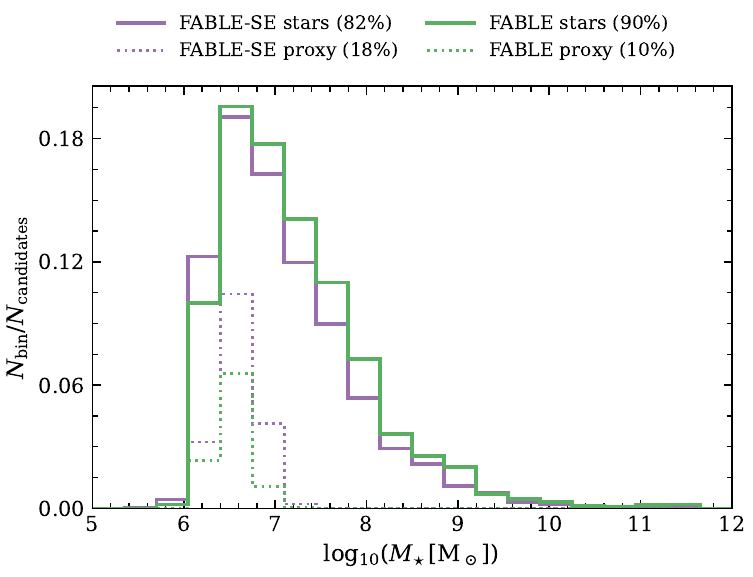}
    \caption{Distributions of stellar mass associated with gas-selected stellar candidates in \textsc{Fable}-SE (purple) and \textsc{Fable} (green). Solid lines show measured stellar masses for clumps containing stellar particles. Dotted lines show estimated stellar masses for starless candidates satisfying our dense-gas proxy, assuming $M_{\star,\mathrm{est}}=0.05\,M_{\mathrm{gas}}$. Both components are normalised to the total number of plotted stellar-candidate detections in each simulation. These masses refer to gas-selected clumps and need not correspond to individual stellar clusters.}
    \label{fig:stellarprop}
\end{figure}

\subsection{Stellar candidates} \label{sec:stellar}
We now examine the stellar content of the stellar candidates defined in Section~\ref{sec:stellar_candidates}. These are gas-selected clumps that either contain stellar particles or, if starless, satisfy our dense-gas proxy. The term `stellar candidate' does not imply that the clump corresponds to a single, resolved stellar cluster. A clump may contain multiple stellar concentrations, potentially including unresolved clusters, as well as diffuse stellar material. We therefore interpret the measured stellar mass as the total stellar mass associated with the gas-selected clump.

Figure~\ref{fig:stellarprop} shows the distributions of measured stellar masses for candidates containing stars and estimated stellar masses for starless candidates, assuming $M_{\star,\mathrm{est}}=\epsilon_\star M_{\mathrm{gas}}$ with $\epsilon_\star=0.05$. The majority of candidate detections contain stars, accounting for $82$ per cent in \textsc{Fable}-SE and $90$ per cent in \textsc{Fable}. Both simulations show broadly similar mass distributions, including when considering only candidates containing stellar particles. Most candidates have measured or estimated stellar masses of $M_\star \sim 10^{6}$--$10^{7.5}\,\mathrm{M}_\odot$, with a tail to higher masses. For starless candidates, these estimates remain conditional on the assumed conversion efficiency and do not establish their future stellar content.

We therefore focus here on the associated stellar mass, without interpreting clump-scale radii and mean surface densities as measurements of individual cluster structure.\footnote{Tests of the sensitivity of the measured stellar properties to the clump-finding parameters are presented in Appendix~\ref{app:clump_finder} (Table~\ref{tab:stellar_finder_sensitivity}). In these tests, median stellar masses vary less than stellar radii and surface densities.} Resolving any constituent stellar systems requires higher-resolution simulations with an appropriate treatment of cold gas and star formation. The subsequent radial analysis follows the gas-clump tracks associated with this same stellar-candidate population.

\subsection{Radial migration} \label{sec:radial_migration}
The previous sections showed that \textsc{Fable}-SE produces more metal-enriched clumps, while the distributions of resolved gas masses, gas sizes and associated stellar masses remain broadly similar to \textsc{Fable}. We now investigate whether the two feedback prescriptions also lead to differences in the spatial distribution and radial kinematics of these systems.

\begin{figure*}
    \centering
        \includegraphics[ trim={0cm 0cm 0cm 0cm}, clip, width=0.99\textwidth, keepaspectratio]{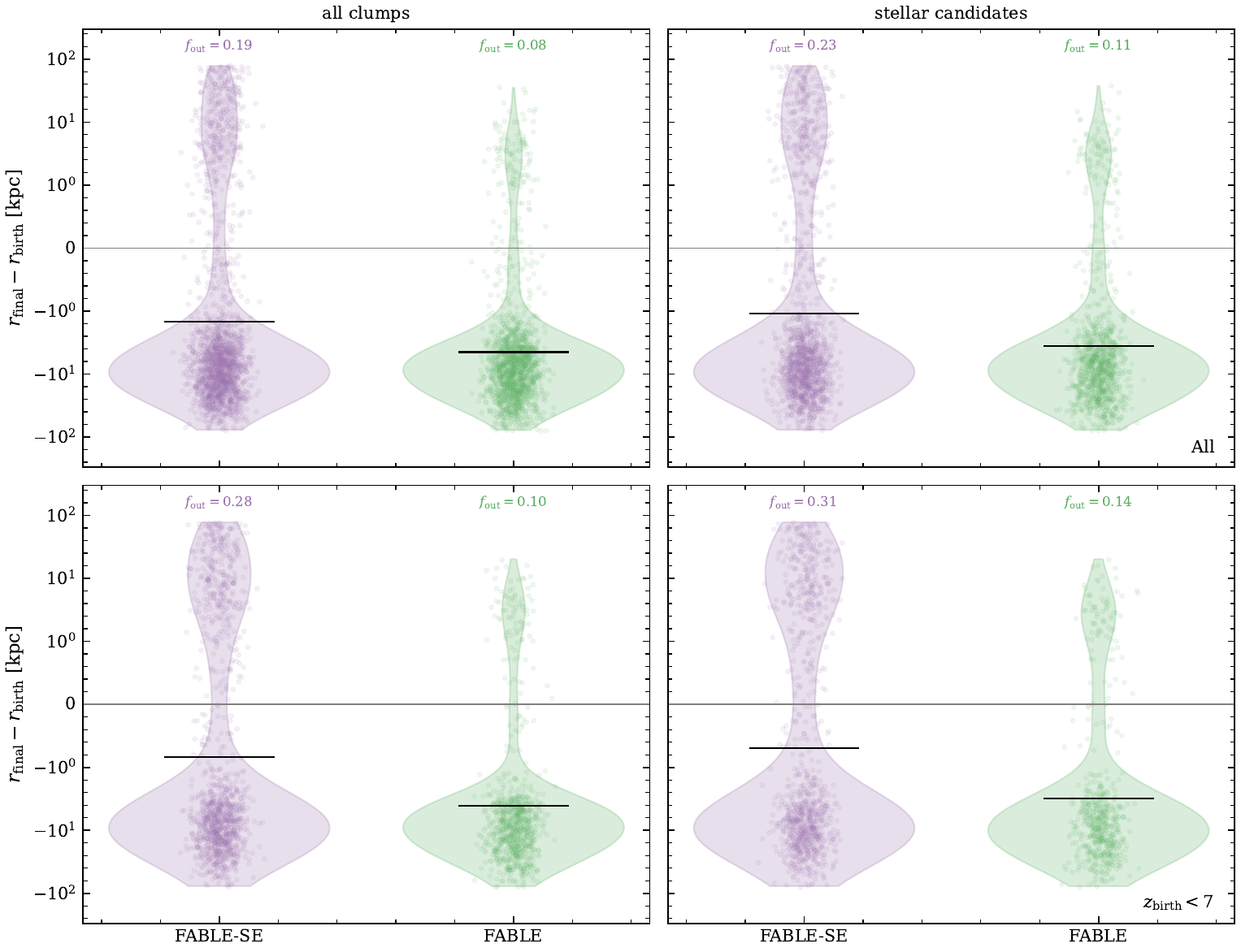}
    \caption{Distribution of net radial migration, $r_{\rm final}-r_{\rm birth}$, for clumps tracked for at least two snapshots. The left panels show all gas clumps, while the right panels show the stellar-candidate sample. The top panels include all birth redshifts, while the bottom panels are restricted to $z_{\rm birth}<7$. Violin plots and overplotted points show the distributions for \textsc{Fable}-SE and \textsc{Fable}, with the horizontal black line marking the mean. Positive values correspond to net outward migration and negative values to net inward migration. The $f_{\rm out}$ values give the fraction of clumps with $r_{\rm final}-r_{\rm birth}>0$. While inward migration dominates in both simulations, \textsc{Fable}-SE has a larger outward-moving population, particularly among stellar candidates born at $z_{\rm birth}<7$ (31 per cent compared to 14 per cent in \textsc{Fable}).}
    \label{fig:violin_migration}
\end{figure*}

\begin{figure*}
    \centering
        \includegraphics[ trim={0cm 0cm 0cm 0cm}, clip, width=1.02\textwidth, keepaspectratio]{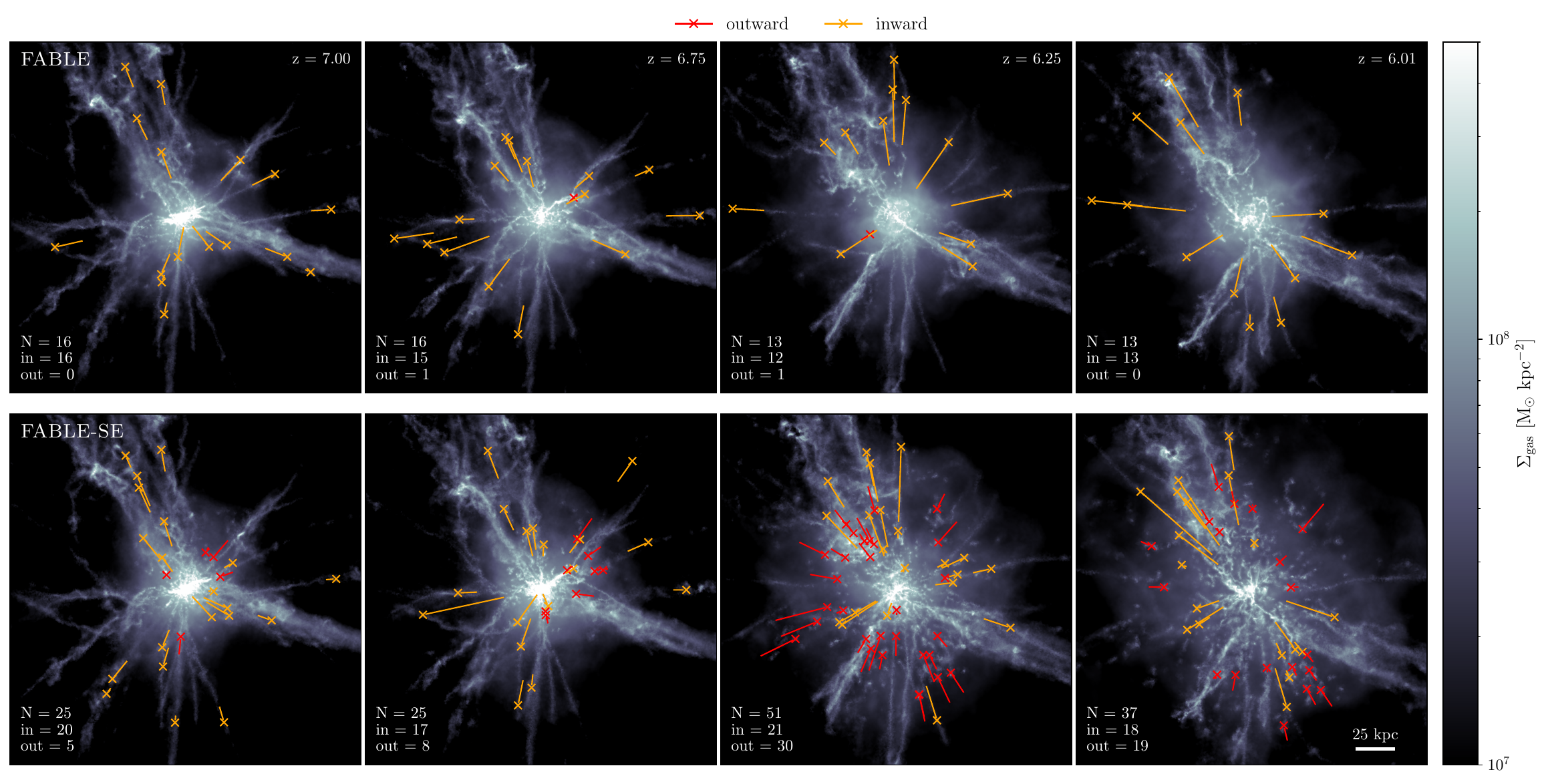}
    \caption{Gas surface density maps centred on the main galaxy for the \textsc{Fable} (top) and \textsc{Fable}-SE (bottom) simulations at different redshifts. Crosses mark stellar candidates, while line segments indicate their subsequent radial displacement over approximately 40 Myr, measured relative to the galaxy centre. Inward-moving systems are shown in orange and outward-moving systems in red. While inward migration dominates in \textsc{Fable}, \textsc{Fable}-SE shows a larger population of outward-moving clumps, particularly at lower redshifts during the AGN-driven blow-out phase.}
    \label{fig:sim}
\end{figure*}

For each tracked clump we define the net radial displacement as
\begin{equation}
    \Delta r = r_{\rm final} - r_{\rm birth}\,,
\end{equation}

\noindent where $r_{\rm birth}$\footnote{Throughout this work, ``birth'' refers to the first snapshot in which a clump is identified within the selected CGM region, and does not necessarily correspond to its physical formation time or location.} is the distance from the central galaxy when the clump first enters the catalogue, and $r_{\rm final}$ is its distance at the last snapshot in which it is identified. Negative values therefore correspond to net inward migration, while positive values indicate that the clump has moved outwards. Since individual tracks span different numbers of snapshots, these displacements represent net radial changes over each clump's identified lifetime, rather than migration over a fixed timescale. A positive displacement therefore does not necessarily imply continuous outward motion.

Figure~\ref{fig:violin_migration} shows the distribution of $\Delta r$ for all gas-selected clumps (left) and for the stellar candidates (right). The top row shows the full sample, while the bottom row includes only clumps born at $z_{\rm birth}<7$, corresponding to the epoch of the strong AGN-driven blow-out phase in \textsc{Fable}-SE.  During this period, powerful outflows disrupt the inner halo and redistribute metal-enriched gas into the CGM, coinciding with the rapid increase in clump abundance discussed previously \citep[see][]{Tortora_et_al_2026}. Separating the sample at $z_{\rm birth}<7$ therefore selects clumps first identified during the disturbed blow-out phase in \textsc{Fable}-SE.

Across all samples shown in Figure~\ref{fig:violin_migration}, between 69 and 92 per cent of clumps have negative $\Delta r$, showing that inward migration remains the dominant behaviour in both simulations. This is expected for dense structures forming in the halo, where gravity and filamentary accretion flows pull material towards the central galaxy.

\begin{figure}
    \centering
        \includegraphics[ trim={0cm 0cm 0cm 0cm}, clip, width=0.48\textwidth, keepaspectratio]{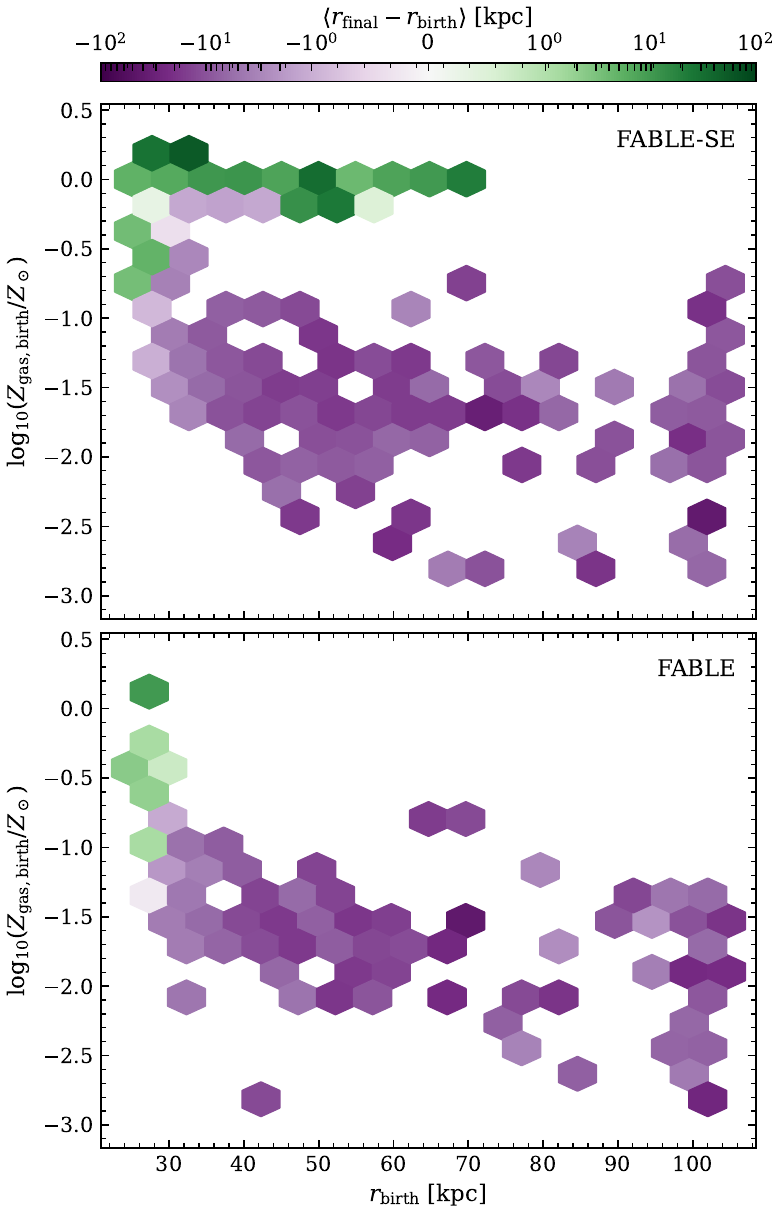}
    \caption{Birth gas metallicity, $\log_{10}(Z_{\rm gas,birth}/Z_\odot)$, as a function of birth radius, $r_{\rm birth}$, for the stellar candidates sample in the \textsc{Fable}-SE and \textsc{Fable} runs. Hexagons are colour-coded by the median net radial migration, $r_{\rm final}-r_{\rm birth}$, in each bin. Only clumps with lifetimes of at least two snapshots are included.}
    \label{fig:dr_metal}
\end{figure}

The two feedback prescriptions nevertheless show a clear difference. \textsc{Fable}-SE has a larger positive tail, with a factor $\approx 2$ higher fraction of outward-moving clumps than \textsc{Fable}. For the full clump sample, the outward fraction is $f_{\rm out}=0.19$ in \textsc{Fable}-SE compared to $0.08$ in \textsc{Fable}. The same trend is present for the stellar candidates, with $f_{\rm out}=0.23$ and $0.11$, respectively. When restricting to objects born at $z_{\rm birth}<7$, after the blow-out event, the difference becomes even stronger: $f_{\rm out}=0.31$ for \textsc{Fable}-SE stellar candidates, compared to $0.14$ in \textsc{Fable}.

Thus, while most clumps still move inwards, the enhanced early BH growth and stronger AGN-driven outflows in Fable-SE are associated with a larger population of outward-moving systems. This suggests that feedback not only changes how many dense clumps form in the CGM, but also changes their subsequent radial evolution. In the rest of this section, we focus on stellar candidates, as these are the systems most directly connected to star formation, either already containing stars or satisfying our dense gas proxy. We further restrict the sample to clumps first identified at $z_{\rm birth}<7$, allowing us to compare their radial evolution during the strong AGN-driven blow-out phase in \textsc{Fable}-SE.

The difference in radial displacement is also visible directly in the gas maps. Figure~\ref{fig:sim} shows gas surface density projections centred on the main galaxy at four different redshifts, with the stellar candidates overplotted. Each cross marks the position of a candidate, while the line segment indicates its subsequent radial motion over $\Delta$~snapshot~$=4$, approximately 40 Myr. Orange lines show inward-moving systems and red lines show outward-moving systems.

The visual difference between the two feedback prescriptions is striking. In \textsc{Fable}, most stellar candidates are associated with filamentary inflow and move inwards towards the central galaxy. Outward-moving systems are rare, and the population remains dominated by infall at all redshifts shown. In \textsc{Fable}-SE, the picture is different. The stellar candidates are more numerous and a much larger fraction move outwards, especially at lower redshift. Many of these systems lie close to the central galaxy or along disturbed gas structures, suggesting that they are connected to feedback-driven motions through the CGM.

Figure~\ref{fig:sim} therefore gives the spatial counterpart to the $\Delta r$ distributions in Figure~\ref{fig:violin_migration}, further showing that the main difference is not simply that \textsc{Fable}-SE forms more stellar candidates, but that many of them are embedded in a more dynamically disturbed CGM. \textsc{Fable}-SE produces a population that is more often pushed outwards, while \textsc{Fable} remains dominated by inward migration along large-scale structure.

Figure~\ref{fig:dr_metal} connects this radial behaviour to the gas metallicity at birth. In both simulations, low-metallicity candidates are mostly associated with negative $\Delta r$, as expected for relatively pristine gas moving inwards along the CGM. The clearest difference appears at higher metallicity. In \textsc{Fable}-SE, metal-rich candidates born at small and intermediate radii are preferentially associated with positive or only weakly negative radial displacements. In \textsc{Fable}, the same region is much less populated and remains dominated by inward-moving systems.

As an independent kinematic check, we examine the instantaneous radial velocities of stellar-candidate detections at $z<7$, without requiring reconstructed track histories. We define positive radial velocity as motion away from the central galaxy. We find that 35.2 per cent of the detections in \textsc{Fable}-SE have positive radial velocities, compared to 12.4 per cent in \textsc{Fable}. Since individual structures may be identified in multiple snapshots, these fractions refer to detections rather than statistically independent clumps. This independently supports a larger population of outward-moving dense gas structures in \textsc{Fable}-SE, consistent with the differences found from the tracking analysis.

Together, these figures show that \textsc{Fable}-SE has a larger population of outward-moving clumps, which are also associated with metal-enriched gas. For stellar candidates born at $z_{\rm birth}<7$, 31 per cent move outwards in \textsc{Fable}-SE, compared to 14 per cent in \textsc{Fable}
\footnote{Figures~\ref{fig:violin_migration} and \ref{fig:dr_metal} include clumps tracked for at least two snapshots. Restricting the stellar candidates born at $z_{\rm birth}<7$ to at least four snapshots gives outward-moving fractions of 40 and 5 per cent in \textsc{Fable}-SE and \textsc{Fable}, respectively.}. This indicates that the stronger AGN-driven blow-out phase in \textsc{Fable}-SE changes not only the metal enrichment of the CGM, but also the spatial distribution and radial evolution of its dense gas structures. 

\begin{figure*}
    \centering
        \includegraphics[ trim={0cm 0cm 0cm 0cm}, clip, width=0.98\textwidth, keepaspectratio]{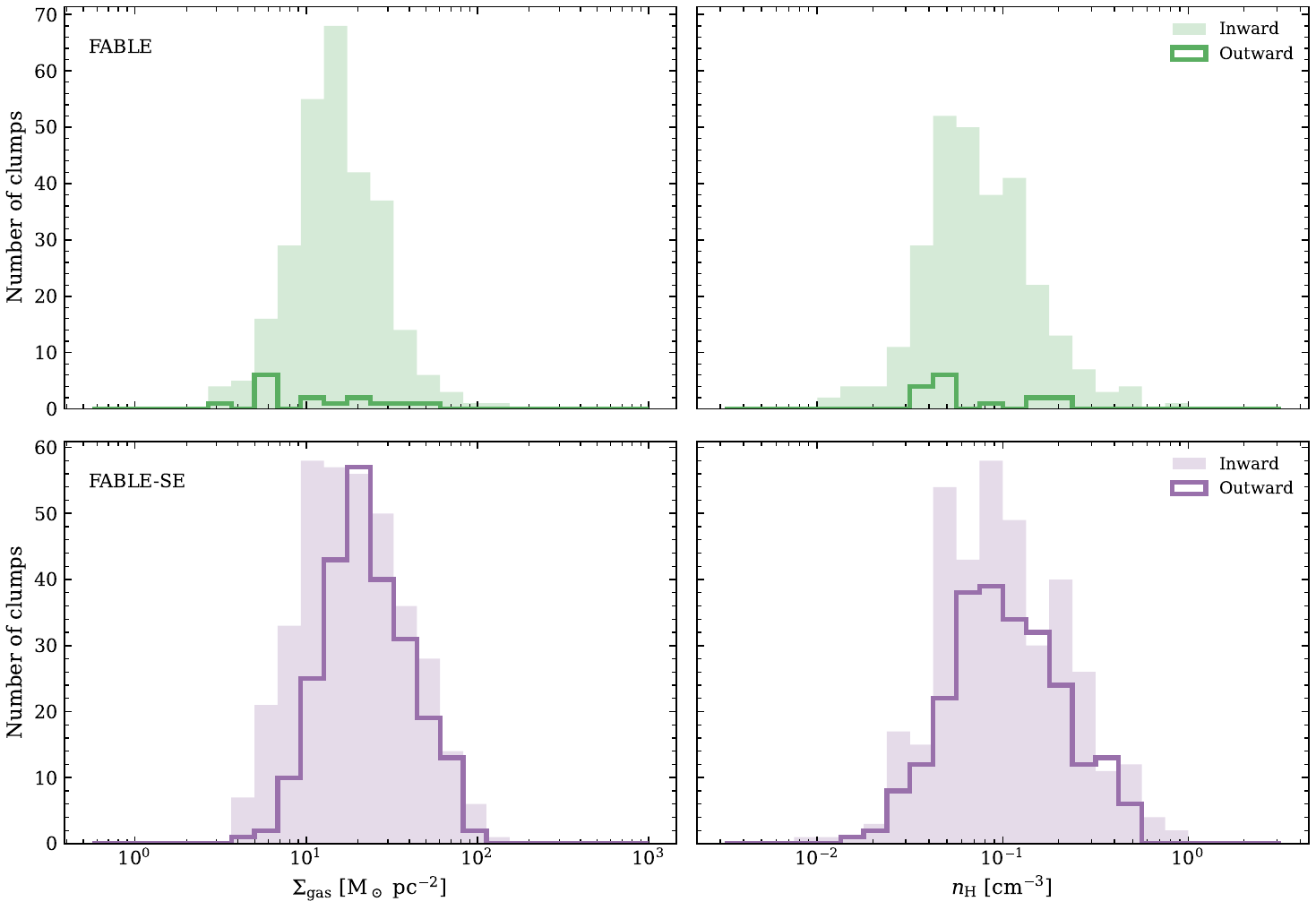}
    \caption{Distributions of gas surface density, $\Sigma_{\rm gas}$, and volume-weighted hydrogen number density, $n_{\rm H}$, for inward- and outward-moving stellar candidates in \textsc{Fable} (top) and \textsc{Fable}-SE (bottom). Filled histograms show inward-moving candidates, while solid outlines show outward-moving candidates. Gas properties are measured at the final identified snapshot. The histograms show the number of clumps per bin, without normalisation. Only candidates born at $z_{\rm birth}<7$ and tracked for at least four snapshots are included. The samples contain 281 inward- and 15 outward-moving candidates in \textsc{Fable}, compared to 367 and 243 in \textsc{Fable}-SE, respectively.}
    \label{fig:gasdensity}
\end{figure*}

\subsubsection{Clumps and satellite galaxies} \label{sec:clumpsat}
As our clump identification is based on gas overdensities rather than gravitationally bound structures, some identified clumps may be associated with resolved satellite galaxies. Such structures are included in our definition of CGM clumps. Nevertheless, we quantify their contribution by cross-matching the clump positions with the nearest non-central subhalo in the main FoF group, considering only subhaloes with total masses above $10^7$ M$_\odot$. We adopt a matching radius of twice the larger of the subhalo's gas and stellar half-mass radii, with a minimum of 5 kpc and a maximum of 25 kpc. A clump is classified as satellite associated if its centre lies within this radius in at least one snapshot along its track. For stellar candidates born at $z_{\rm birth}\geq7$, approximately 35 per cent are associated with satellites in both simulations. At $z_{\rm birth}<7$, satellite associations are less common in \textsc{Fable}-SE, particularly among outward-moving candidates ($\sim 10$ per cent compared to $\sim 50$ per cent in \textsc{Fable}). 

For candidates born at $z_{\rm birth}<7$, the number of satellite-associated systems is similar between the simulations. In contrast, the number of candidates not associated with resolved satellites is substantially higher in \textsc{Fable}-SE (485 compared to 180). This shows that the additional clump population in \textsc{Fable}-SE is not primarily associated with resolved satellites. Instead, it is consistent with dense CGM structures embedded in the disturbed and enriched halo produced during the stronger AGN-driven blow-out phase.

\subsubsection{Gas properties of migrating clumps} \label{sec:SF}

\begin{figure*}
    \centering
        \includegraphics[ trim={0cm 0cm 0cm 0cm}, clip, width=1.02\textwidth, keepaspectratio]{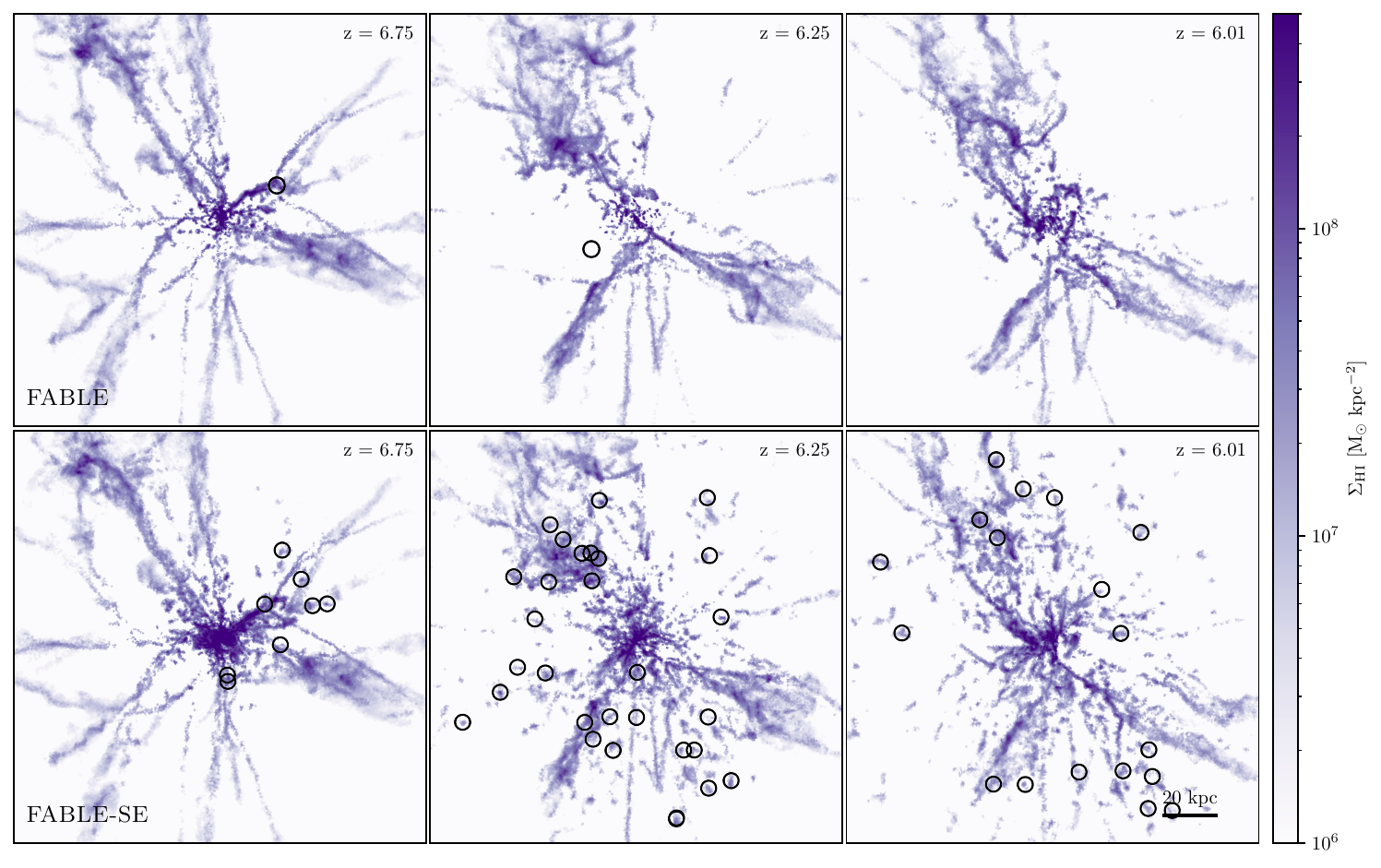}
    \caption{HI gas surface density maps centred on the main galaxy for the \textsc{Fable} (top) and \textsc{Fable}-SE (bottom) simulations, shown at different redshifts. Circles mark outward-moving stellar candidate clumps with lifetimes of at least four snapshots.}
    \label{fig:HImaps}
\end{figure*}

The previous sections showed that \textsc{Fable}-SE produces a larger population of outward-moving stellar candidates, particularly during the strong blow-out phase at $z \lesssim 7$. We now examine whether these systems remain associated with relatively dense gas as they move through the CGM.

Figure~\ref{fig:gasdensity} compares the gas surface density, $\Sigma_{\rm gas}$, and hydrogen number density, $n_{\rm H}$, of inward- and outward-moving stellar candidates. In \textsc{Fable}-SE, the outward-moving population has a slightly higher median gas surface density than the inward-moving population, with values of $21.4$ and $17.8$~M$_\odot$~pc$^{-2}$, respectively. Their median hydrogen number densities are very similar, at $0.100$ and $0.094$~cm$^{-3}$. These differences remain modest when restricting the comparison to clumps containing stellar particles.

In contrast, outward-moving candidates in \textsc{Fable} have lower median gas surface densities and hydrogen number densities than their inward-moving counterparts. Their median $\Sigma_{\rm gas}$ decreases from $14.8$ to $9.6$~M$_\odot$~pc$^{-2}$, while $n_{\rm H}$ decreases from $0.069$ to $0.045$~cm$^{-3}$. However, the outward-moving \textsc{Fable} sample contains only 15 candidates, so these differences should be interpreted cautiously.

Overall, the outward-moving population in \textsc{Fable}-SE is associated with relatively dense gas at the final identified snapshot. Together with their enhanced metallicities and outward migration, this is consistent with the redistribution of dense, enriched gas during the AGN-driven blow-out phase. However, as our stellar-candidate selection also includes dense starless gas clumps, these density measurements do not establish ongoing star formation within the identified structures.

Figure~\ref{fig:HImaps} further illustrates the different CGM environments produced in the two simulations. The figure shows HI surface density maps for the \textsc{Fable} and \textsc{Fable}-SE simulations, with outward-moving stellar candidates indicated by the circles. In \textsc{Fable}, the HI distribution remains relatively calm and filamentary, with most of the dense neutral gas tracing coherent inflowing structures connected to the cosmic web. Outward-moving systems are rare, with most remaining associated with the large-scale filamentary inflow. In contrast, the CGM in \textsc{Fable}-SE appears significantly more disturbed during the blow-out phase \citep[see also][]{Tortora_et_al_2026}. Dense HI-rich gas is distributed throughout the halo, producing a more irregular and fragmented neutral gas morphology around the central galaxy. A substantially larger population of outward-moving stellar candidates is embedded within these disturbed HI-rich structures surrounding the galaxy. 

\begin{table*}
\centering
\caption{Median gas properties of inward- and outward-moving stellar candidates born at $z_{\rm birth}<7$. Only candidates tracked for at least four snapshots are included. The cold gas fraction, $f_{\rm cold}$, includes all gas on the effective equation of state (eEoS), together with non-star-forming gas at $T<5\times10^4$ K. Gas on the eEoS is identified through non-zero star formation rates and is assumed to be cold irrespective of its effective temperature. This accounts for the unresolved multiphase nature of star-forming gas. Outward-moving \textsc{Fable}-SE candidates have substantially higher cold gas fractions than their counterparts in \textsc{Fable}.}
\label{tab:temperature_properties}
\begin{tabular}{llccccc}
\hline
Simulation & Direction & $N$ & $T_{\rm gas}$ [K] & $f_{\rm cold}$ & $f_{\rm cold}>0.5$ & $\nabla\cdot v$ [km s$^{-1}$ kpc$^{-1}$] \\
\hline
\textsc{Fable}    & Inward  & 281 & $4.8\times10^5$ & 0.66 & 0.65 & $-91$ \\
\textsc{Fable}    & Outward & 15  & $2.0\times10^6$ & 0.23 & 0.40 & $-46$ \\
\textsc{Fable}-SE & Inward  & 367 & $4.0\times10^5$ & 0.72 & 0.67 & $-144$ \\
\textsc{Fable}-SE & Outward & 243 & $8.9\times10^4$ & 0.92 & 0.96 & $-156$ \\
\hline
\end{tabular}
\end{table*}

We further quantify the physical state of the inward- and outward-moving stellar candidates using the gas properties measured within the clumps. We define the cold gas fraction as the fraction of gas mass either on the effective equation of state (eEoS) or, for non-star-forming gas, at temperatures $T<5\times10^4$~K. For gas on the eEoS, we assume that all of its mass contributes to the cold component, following the unresolved multiphase treatment of the star-forming ISM. This is a modelling assumption rather than a direct measurement of the resolved cold gas phase.

Table~\ref{tab:temperature_properties} shows that the outward-moving systems in \textsc{Fable}-SE contain substantially more cold gas than their counterparts in \textsc{Fable}. For candidates born at $z_{\rm birth}<7$, the median mass-weighted gas temperature of the outward-moving population is $8.9\times10^4\,{\rm K}$ in \textsc{Fable}-SE, compared to $2.0\times10^6\,{\rm K}$ in \textsc{Fable}. However, these temperatures include gas on the effective equation of state (eEoS), for which the assigned temperature does not represent a resolved physical thermal phase. Following the unresolved multiphase treatment of star-forming gas, we therefore classify all gas on the eEoS as cold, regardless of its effective temperature, together with non-star-forming gas at $T<5\times10^4$~K. With this definition, the median cold gas fraction is $f_{\rm cold}=0.92$ in \textsc{Fable}-SE, compared to $0.23$ in \textsc{Fable}. Almost all outward-moving \textsc{Fable}-SE candidates ($96$ per cent) have more than half of their gas classified as cold, compared to only $40$ per cent in \textsc{Fable}. These results remain largely unchanged compared to applying the temperature threshold to all gas cells. To assess whether this difference is driven entirely by the eEoS treatment, we also calculate the fraction of non-star-forming gas below $5\times10^4$~K, relative to the total gas mass. This gives a median of $0.38$ in \textsc{Fable}-SE, compared to $0.00$ in \textsc{Fable}. The difference between the simulations is therefore not solely driven by the treatment of gas on the eEoS. Combined with the gas-density distributions in Figure~\ref{fig:gasdensity}, these results suggest that the outward-moving population in \textsc{Fable}-SE remains associated with relatively cool, dense gas while moving through the disturbed CGM. However, we cannot determine whether these structures continue to form stars.

The $T-n_{\rm H}$ plane shown in Appendix~\ref{app:tnh_plane} further illustrates these differences. The outward-moving \textsc{Fable}-SE systems span a broad range of temperatures and densities, including relatively cool and dense gas, rather than being confined to the hottest gas in the sample. Their gas motions are also locally converging, with a median velocity divergence of $-156$ km s$^{-1}$ kpc$^{-1}$. These converging motions are consistent with local gas compression, although they do not establish how the clumps form or are maintained. To investigate this further, we compare their effective gas pressure with that of the surrounding CGM. The outward-moving \textsc{Fable}-SE candidates have median internal-to-surrounding pressure ratios of 0.24 and 0.25 for mass- and volume-weighted pressures, respectively, using a shell at $1.5$--$4R_{\rm ap}$ ($N=241$). A narrower shell at $1$--$1.5R_{\rm ap}$ gives ratios of 0.69 and 0.53 ($N=206$). These results show that locally converging gas motions do not necessarily correspond to higher internal effective pressure. Cooling and gas accretion may contribute to this behaviour, but we cannot determine their role from these measurements alone.

We further examine whether these locally converging gas motions are associated with gravitationally bound structures on the resolved scales. For the outward-moving stellar candidates born at $z_{\rm birth}<7$, we find median virial parameters of $\alpha_{\rm vir}\sim7.8$ in \textsc{Fable} and $\sim6.7$ in \textsc{Fable}-SE. These relatively large values suggest that the gas structures are not strongly self-gravitating on the resolved scales, although the estimate does not account for the full gravitational potential or the unresolved internal structure. The median convergence ratios are $C\sim1.4$--$1.5$ in both simulations, indicating that converging radial gas motions are present but do not strongly distinguish the two populations.

Together, these results suggest that the outward-moving population in \textsc{Fable}-SE represents extended, relatively cool and dense gas structures embedded within the disturbed CGM. Although these systems are not resolved as compact stellar clusters, they may contain smaller, unresolved star-forming regions. However, our measurements do not establish ongoing star formation or the long-term survival of individual clumps.

Some of these structures also change their location relative to the virial radius during their tracked evolution. In \textsc{Fable}-SE, 52 stellar-candidate tracks (8.5 per cent) are first identified inside $R_{\rm vir}$ and last identified outside it, compared to none in \textsc{Fable}. This difference remains when adopting stricter tracking criteria (see Appendix~\ref{app:tracking}).

\section{Discussion and conclusions}\label{sec:disc}
In this work, we compare two cosmological zoom-in simulations of a massive high-redshift quasar-host halo, \textsc{Fable} and \textsc{Fable}-SE, evolved with \textsc{Arepo}. Both simulations follow the same halo, reaching a virial mass of approximately $7\times10^{12}\,{\rm M_\odot}$ at $z=6$, and use the same prescriptions for gas cooling, star formation, and stellar feedback. While \textsc{Fable} adopts Eddington-limited black hole accretion, \textsc{Fable}-SE allows earlier and mildly super-Eddington black hole growth, resulting in stronger early AGN feedback.

The high-redshift CGM is expected to contain a complex mixture of inflowing filamentary gas, recycled material, mergers, and feedback-driven outflows coexisting within the same halo \citep[e.g.][]{dekel:2009aa, angles:2017aa, ginolfi:2020aa, costa:2022aa}. Our results show that the stronger early feedback associated with enhanced black hole growth changes the distribution and radial motion of dense gas structures within the CGM. While the overall clump populations in \textsc{Fable} and \textsc{Fable}-SE have broadly similar resolved gas masses and sizes, as well as associated stellar masses, \textsc{Fable}-SE develops an additional population of metal-enriched and outward-moving systems during the strong AGN-driven blow-out phase at $z \lesssim 7$ \citep[for more details on this blow-out phase see][]{Tortora_et_al_2026}.

This suggests that the stronger AGN-driven blow-out phase in \textsc{Fable}-SE does not simply remove cold gas from the halo, but is also associated with more dense, enriched material throughout the CGM and a more multiphase and dynamically disturbed environment \citep[e.g.][]{Costa:2015aa, costa:2018aa, costa:2020aa, ward:2024aa} partly through outward migration. Importantly, inward migration remains common in \textsc{Fable}-SE, despite the stronger feedback. We instead find an additional population of metal-enriched, outward-moving clumps consistent with the redistribution of dense gas by AGN-driven outflows within the halo \citep[e.g.][]{angles:2017aa}. Inward migration through filamentary accretion nevertheless remains the dominant evolutionary pathway in both simulations, consistent with the broader picture of cosmological gas feeding at high redshift \citep[e.g.][]{dekel:2009aa, keres:2005aa, vanDonkelaar:2023aa, Donkelaar:2026aa}.

The connection between radial migration, metallicity, and the disturbed HI-rich CGM morphology further supports this interpretation. Low-metallicity systems in both simulations are predominantly associated with inward migration, consistent with relatively pristine inflowing gas. In contrast, the outward-moving population in \textsc{Fable}-SE is additionally preferentially embedded within disturbed HI-rich structures, see Figure~\ref{fig:HImaps}, suggesting that these systems are closely connected to gas displaced through the halo during the blow-out phase.

Together, these results suggest that dense circumgalactic clumps can follow different evolutionary pathways, even within the same high-redshift halo. Some systems remain connected to filamentary inflow and continue migrating towards the central galaxy, while others show outward migration associated with AGN-driven feedback, including crossings of the virial radius that are only found in \textsc{Fable}-SE. This behaviour is particularly interesting in the context of recent \textit{JWST} observations suggesting that compact stellar systems and clustered star formation may already be common in dense gas-rich environments during cosmic dawn \citep[e.g.][]{Mowla:2022aa, Adamo:2024aa, Messa:2025aa, Whitaker:2025aa}.

The presence of dense gas in the CGM has also been studied in other cosmological simulations. For example, \citet{Suresh:2019aa} found that galactic winds can increase the amount of cold gas in the halo, with much of this gas cooling after interacting with the surrounding hot CGM. Similarly, \citet{Nelson:2020aa} found a large population of cold gas clouds in TNG50, forming through local gas overdensities and subsequent cooling. Dense structures can also form through the fragmentation of accreting gas filaments, potentially leading to the formation of compact stellar systems \citep[e.g.][]{Mandelker:2018aa,vanDonkelaar:2023aa,Donkelaar:2026aa}. While inward migration dominates in both our simulations, \textsc{Fable}-SE also produces a larger population of outward-moving, metal-enriched clumps during the AGN-driven blow-out phase. Our results therefore suggest that both gas accretion and feedback can contribute to the population of dense structures in the CGM, although their relative importance will likely depend on the halo environment and feedback model.

The formation and survival of these structures also depend on their interaction with the surrounding gas. Previous simulations have shown that clump survival is sensitive to feedback and the local gas conditions \citep[e.g.][]{Mandelker:2017aa,Mandelker:2020aa}. In particular, radiative cooling can allow cold streams to survive hydrodynamical instabilities, while cooling, fragmentation and coagulation can further alter the structure of cold gas within the CGM \citep[see also][]{Yao:2025aa,Yao:2026aa}. These processes may also be important for the clumps identified in our simulations. However, given our limited resolution, we cannot follow their internal fragmentation or determine whether the outward-moving clumps in \textsc{Fable}-SE were already dense when transported outwards or formed through cooling and condensation within the disturbed CGM.

\subsection{Fleeing Dense Gas Structures} \label{sec:FleeingSFR}
The outward-moving population in \textsc{Fable}-SE appears qualitatively different from the more inflow-dominated systems discussed in the previous section. Their outward migration, elevated gas-phase metallicities and association with relatively dense, HI-rich gas are consistent with the displacement of dense material through the halo during the AGN-driven blow-out phase \citep[e.g.][]{ishibashi:2012aa, zubovas:2014aa, Costa:2015aa}. However, our current analysis cannot distinguish whether the gas was already dense when transported outwards or became dense through cooling and condensation within the disturbed CGM. These systems may contain sites of recent or ongoing star formation, but their resolved gas properties alone cannot establish this.

This behaviour is less pronounced in \textsc{Fable}, where outward-moving systems are rarer, generally hotter, and have lower cold gas fractions and lower median gas densities. However, the outward-moving \textsc{Fable} sample is small. Together with the disturbed HI morphology surrounding the galaxy during the blow-out phase in \textsc{Fable}-SE, this suggests that the stronger early feedback is associated with the redistribution of relatively cool, dense and enriched gas through the CGM. Such a dense and clumpy CGM is also consistent with recent constraints from quasar Ly$\alpha$ nebulae at lower redshift, which suggest the presence of gas reaching densities of at least $\sim10\,{\rm cm^{-3}}$ and substantially more small-scale density structure than predicted by most current simulations \citep{Lazeyras:2026aa}. The higher cold gas fractions in the outward-moving population of \textsc{Fable}-SE further suggest that these systems contain substantial amounts of relatively cool gas at their last identified snapshots, although this does not establish the survival time of individual gas structures.

At the same time, the outward-moving systems should not be interpreted as fully bound stellar clusters on the resolved kiloparsec scales. As shown in Section~3.3.2, their relatively large virial parameters suggest that the gas structures are not strongly self-gravitating on these scales, despite the presence of locally converging gas motions. Interestingly, the outward-moving systems in \textsc{Fable}-SE show more negative velocity divergences than those in \textsc{Fable}, while their convergence ratios remain broadly similar. These diagnostics measure different aspects of the gas motions: the velocity divergence describes local spatial convergence, whereas $C$ compares the mass-weighted inward and outward radial motions relative to the clump centre. A more negative velocity divergence therefore does not necessarily imply a larger convergence ratio. These measurements also cannot rule out smaller, gravitationally bound structures below the resolution limit of the simulations.

Together, these results suggest that the outward-moving population in \textsc{Fable}-SE represents extended, relatively cool and dense gas structures that may contain unresolved sites of star formation, rather than resolved, gravitationally bound stellar clusters. Whether individual structures survive intact or contain smaller, gravitationally bound stellar systems remains uncertain.

\begin{figure}
    \centering
        \includegraphics[ trim={0cm 0cm 0cm 0cm}, clip, width=0.48\textwidth, keepaspectratio]{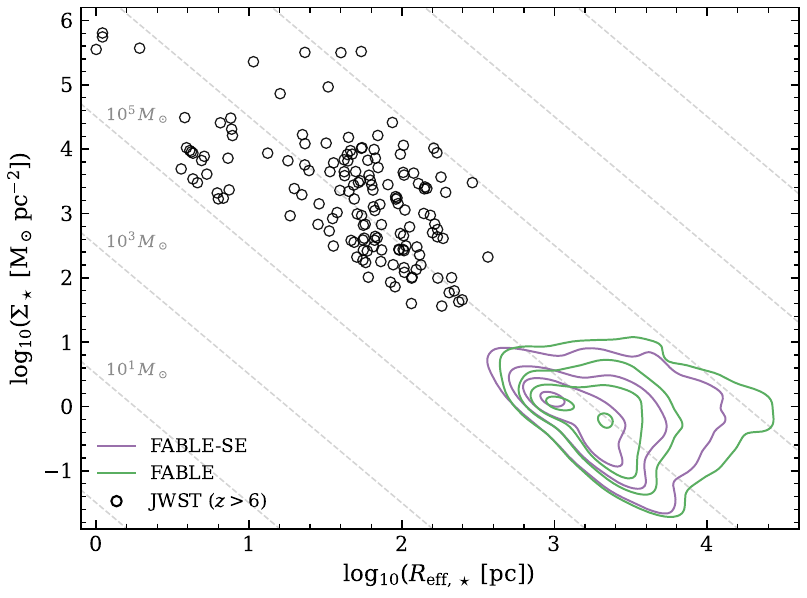}
    \caption{Stellar surface density, $\Sigma_{\star}$, as a function of effective radius, $R_{\rm eff,\star}$, for the stellar candidates sample. Contours show the distribution of simulated clumps in the \textsc{Fable}-SE (purple) and \textsc{Fable} (green) runs. Open circles show \textit{JWST} clusters at $z > 6$ from \citet{Claeyssens:2023aa, Adamo:2024aa, Fujimoto:2024aa, Mowla:2024aa, Messa:2025aa}. Grey dashed lines indicate constant stellar mass.  }
    \label{fig:JWST}
\end{figure}

\subsection{Are these objects observable?} \label{sec:observe}
The gas-rich, metal-enriched clumps in \textsc{Fable}-SE could contribute to metal absorption systems observed around high-redshift galaxies and quasars \citep[e.g.][]{Davies:2023aa,Sebastian:2024aa,Zou:2024aa}. In particular, the outward-moving population may provide an additional source of enriched gas in the CGM, alongside inflowing and recycled material. Absorption observations could therefore provide a complementary probe of these dense circumgalactic structures, although predicting individual absorption features requires modelling the ionisation state, column densities and projected distribution of the gas.

Where young stellar populations are present, the stellar candidates could be associated with spatially offset rest-frame UV emission around high-redshift galaxies and quasars. However, the masses in Figure~\ref{fig:stellarprop} refer to the total stellar component associated with each gas-selected clump and do not specify the luminosity, size or surface brightness of an individual stellar source. For starless candidates, the estimated masses describe a hypothetical conversion of gas into stars, rather than an existing stellar population. 

Figure~\ref{fig:JWST} compares our stellar candidates with compact stellar systems observed with \textit{JWST}. The simulated candidates typically have larger radii and lower mean stellar surface densities. Our gas-selected clumps may contain several smaller stellar concentrations, whose internal structure remains unresolved. Higher-resolution simulations with a more detailed treatment of cold gas and star formation are needed to investigate whether these clumps host or form compact stellar clusters.

\subsection{Limitations} \label{sec:limit}
\subsubsection{Impact of numerical resolution and clump identification}\label{sec:res}
Beyond the resolution limitations discussed above, it is known that the small-scale structure of the CGM is not converged in galaxy formation simulations \citep[e.g.,][]{Hummels+_2019, vandeVoort+_2019, Ramesh_Nelson_2024}. Larger quantities of cool gas are found to condense out of the CGM and structures on the smallest scales are more abundant \citep[][]{Bennett_Sijacki_2020, Lucchini+_2026, Ramesh+_2026, Almeida+_2026}. The number of clumps identified in this work may thus increase significantly with higher-resolution simulations, while their properties may differ as we model the collapse and evolution of these substructures on sub-kpc scales, which is not captured in this work. 

As discussed in Section~\ref{sec:identifiy_clump},  we test linking lengths of 0.25, 0.5 and 1 kpc, overdensity thresholds of 200 and 500, and minimum gas-cell counts of 16, 24, 50 and 100 (Appendix~\ref{app:clump_finder}). The overdensity threshold has relatively little effect on the measured clump properties, while the linking length has a much stronger impact. Smaller linking lengths identify denser structures with smaller radii, but also substantially reduce the number of retained clumps at 0.25 kpc. This is partly because larger overdensities are separated into smaller groups that no longer satisfy our minimum cell-count criterion. For linking lengths of 0.5--1 kpc, the enhanced metal-rich population in \textsc{Fable}-SE remains present across the tested selections, while at 0.25 kpc this difference is not consistently recovered.

Importantly, our aim is not to resolve individual dense clouds or stellar clusters, but to identify larger overdense gas regions within the CGM. These may contain smaller structures that would only become apparent at higher numerical resolution. This was also an important motivation for our adopted linking length: reducing it at fixed resolution does not necessarily provide a more physical description of the internal clump structure. Resolving whether these gas complexes fragment into smaller, potentially star-forming systems requires higher-resolution simulations and is beyond the scope of this work.

Interestingly, recent comparisons of quasar CGM observations with simulations suggest that the cold CGM may be substantially denser and more clumpy than currently resolved, with higher-resolution simulations recovering more of this structure \citep[][see also \citealt{Escamilla:2026aa}]{Lazeyras:2026aa}. At our resolution, the adopted linking length may connect several neighbouring gas overdensities, potentially along filaments, into a single clump. Observations of high-redshift galaxies also show that the interpretation of a clump depends strongly on spatial resolution. In the SCALES studies, \citet{Kalita:2026aa, Kalita:2026ab} identify smaller structures of $\sim100$--$300$ pc within regions that would be observed as giant clumps on kpc scales. They find that these larger structures may represent either coherent giant clumps containing smaller substructures or associations of individual clumps. Although these studies focus on stellar structures within galaxies rather than gas overdensities in the CGM, they illustrate the importance of resolving the internal structure of a clump. We therefore do not interpret our identified systems as individual compact clouds or stellar clusters, as their internal fragmentation remains unresolved.

The current simulations only allow us to follow the CGM clumps over relatively short timescales. Some clumps remain identifiable for a few hundred Myr, while others can only be tracked for much shorter periods. However, when a clump disappears from the catalogue, we cannot always determine whether it has dispersed, fragmented, or no longer meets our selection criteria. In particular, it remains unclear whether the outward-moving systems stay coherent, fragment into smaller structures, or eventually fall back onto the galaxy. Higher-resolution simulations extending to lower redshifts, together with more detailed tracking, will be needed to determine their longer-term fate.

\subsubsection{Impact of the subgrid physics} \label{sec:ism}
There are several modelling assumptions made in this work that are subject to further discussion. An important caveat is the dependence on the adopted AGN feedback model. The enhanced CGM clump population in \textsc{Fable}-SE is associated with the strong AGN-driven blow-out phase at $z \lesssim 7$, following enhanced early BH growth, during which powerful outflows transport cold, metal-rich gas into the halo \citep[see][]{Tortora_et_al_2026}. While this produces a consistent physical picture connecting AGN feedback and outflows to the abundance and evolution of CGM clumps, the quantitative strength of this effect will depend on the SMBH accretion and feedback prescription. Different implementations may alter both the efficiency of the outflows and the survival of cold gas in the halo \citep[e.g.][]{Roos:2015aa}.
Furthermore, our comparison is limited to one massive halo, and it remains unclear whether the same trends occur across different halo masses and assembly histories.

Another important limitation of this work is that the \textsc{Fable} model was not designed to resolve realistic compact cluster formation or the detailed structure of the cold ISM. Star-forming gas is treated using an effective multiphase subgrid model, where dense gas follows a pressurised equation of state and stars form stochastically once a density threshold is exceeded \citep[see also][]{Springel_Hernquist_2003, Vogelsberger_et_al_2013, Henden:2018aa}. As a result, the simulations do not explicitly resolve the cold, turbulent fragmentation and collapse of molecular clouds on small scales.

Motivated by these limitations, our stellar-candidate sample includes both gas clumps containing stellar particles and starless clumps satisfying the density proxy defined in Section~\ref{sec:stellar_candidates}. Stochastic stellar-particle creation and unresolved gas structure can affect whether a dense clump contains stars at a given snapshot. Including the dense starless subset allows us to examine such gas structures without requiring an existing stellar component, although it does not establish whether they will subsequently form stars.

This is particularly important for the dense CGM structures studied in this work. The combination of limited spatial resolution and the effective ISM model may artificially smooth or stabilise dense gas clouds moving through the halo, while the coupling between AGN feedback and unresolved star-forming gas remains uncertain. The survival, fragmentation, and compactness of the identified systems may therefore differ in simulations with explicit cold ISM physics or resolved cluster formation models. Future higher-resolution simulations with more detailed ISM and star formation prescriptions are needed to investigate star formation within these gas structures and the formation and survival of any compact stellar components.

For this reason, we deliberately refer to these systems as stellar candidates throughout this work, rather than fully resolved stellar clusters. The main results should therefore primarily be interpreted in terms of the large-scale evolution of dense CGM structures during the blow-out phase, rather than the detailed formation of individual compact stellar systems.

At the same time, the broader trends found here, including the disturbed HI morphology, the enhanced population of outward-moving systems, and their connection to dense, metal-enriched gas, likely reflect differences in the large-scale CGM evolution between the two feedback prescriptions.

\subsection{Summary}
We summarise our main findings as follows:

\begin{itemize}
    \item Both simulations produce CGM clumps with similar resolved gas masses and sizes. However, \textsc{Fable}-SE produces up to twice as many clumps during the strong AGN-driven blow-out phase at $z \lesssim 7$.
    
    \item The main difference is in metallicity. At $z<7$, 43 per cent of clump detections in \textsc{Fable}-SE are metal-rich, compared to only 9 per cent in \textsc{Fable}. These metal-rich clumps are also generally smaller and have higher gas surface densities.

    \item Inward migration remains the dominant behaviour in both simulations, consistent with gas accretion along cosmic filaments. However, \textsc{Fable}-SE also produces a larger outward-moving population. For stellar candidates born at $z_{\rm birth}<7$, 31 per cent move outwards, compared to 14 per cent in \textsc{Fable}. Some even move beyond the virial radius.

    \item The outward-moving population in \textsc{Fable}-SE is associated with enriched gas, disturbed HI-rich structures and relatively dense gas environments. Their median gas surface density is slightly higher than that of inward-moving candidates, while their hydrogen number densities are broadly similar. Combined with their outward migration, this is consistent with the redistribution of dense, metal-enriched material during the AGN-driven blow-out phase. However, we cannot determine whether this gas was already dense when transported outwards or became dense within the halo, nor whether individual structures survive or eventually form compact stellar clusters.

\end{itemize}

These results suggest different pathways for dense structures to populate the CGM. Some originate in the cosmic filaments feeding galaxies, while others may be displaced outwards by strong feedback. A clump found far from a galaxy therefore does not necessarily have to form there. Understanding these systems requires following both where they came from and how they evolve afterwards.

%%%%%%%%%%%%%%%%%%%%%%%%%%%%%%%%%%%%%%%%%%%%%%%%%%
\section*{Acknowledgements}
We thank Jake Bennett for carrying out the simulations used in this work. FvD acknowledges support from the Herchel Smith Fellowship at the University of Cambridge. FvD would like to thank Matthew Smith for providing \textsc{vortrace} to create the projection maps. LT acknowledges support from the Science and Technology Facilities Council (STFC) for a PhD studentship (STFC Quota Award - ST/Y509139/1). DS acknowledges support from the Science and Technology Facilities Council (STFC) under grant ST/W000997/1. This work used the DiRAC Data Intensive service (CSD3) at the
University of Cambridge, managed by the University of Cambridge
University Information Services on behalf of the STFC DiRAC HPC
Facility (www.dirac.ac.uk). The DiRAC component of CSD3 at
Cambridge was funded by BEIS, UKRI and STFC capital funding
and STFC operations grants. This work used the DiRAC Memory
Intensive service (Cosma8) at Durham University, managed by the In-
stitute for Computational Cosmology on behalf of the STFC DiRAC
HPC Facility (www.dirac.ac.uk). The DiRAC service at Durham
was funded by BEIS, UKRI and STFC capital funding, Durham Uni-
versity and STFC operations grants. DiRAC is part of the UKRI
Digital Research Infrastructure.

%%%%%%%%%%%%%%%%%%%%%%%%%%%%%%%%%%%%%%%%%%%%%%%%%%
\section*{Data Availability}
The data underlying this article will be shared on reasonable request to the corresponding author.

%%%%%%%%%%%%%%%%%%%% \textsc{Fable}-SES %%%%%%%%%%%%%%%%%%

% The best way to enter \textsc{Fable}-SEs is to use BibTeX:

\bibliographystyle{mnras}
\bibliography{example} % if your bibtex file is called example.bib

\newpage
\appendix

\section{Clump identification and tracking tests}
\label{app:robustness}

Here, we test how the adopted clump-finding parameters and tracking criteria affect our results.

\subsection{Clump identification}
\label{app:clump_finder}

We test linking lengths of 0.25, 0.5 and 1 kpc, overdensity thresholds of 200 and 500, and minimum gas-cell counts of 16, 24, 50 and 100. These tests are performed in both simulations using four snapshots between $z=6.51$ and $z=5.75$.

\begin{figure}
    \centering
    \includegraphics[width=\columnwidth,keepaspectratio]{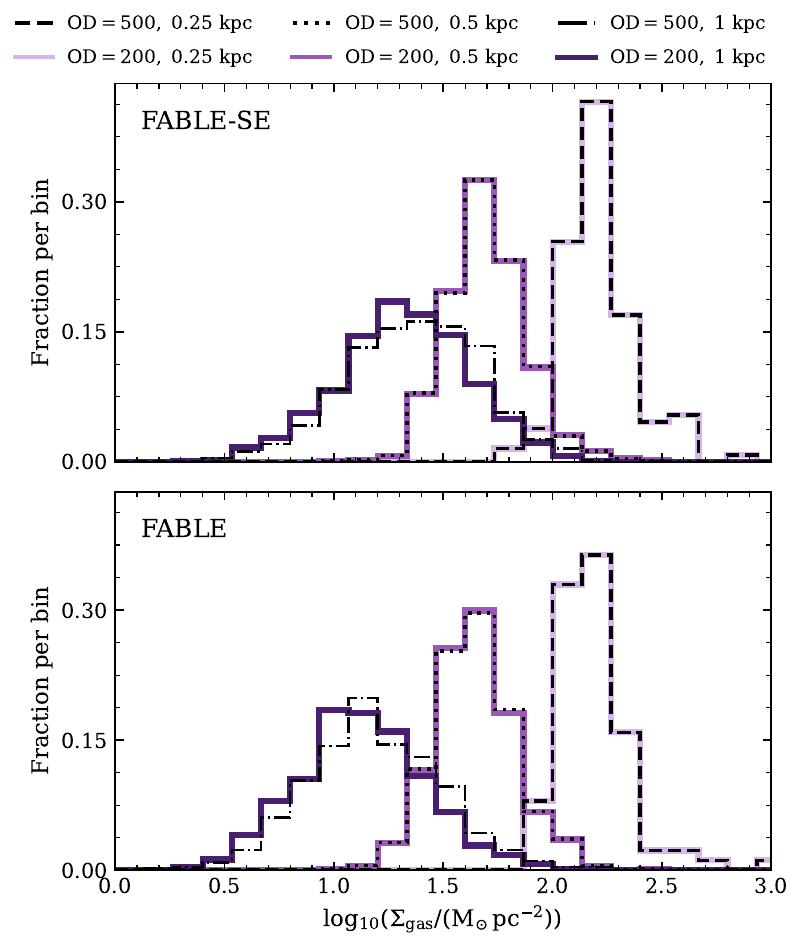}
    \caption{
    Gas surface-density distributions of stellar candidates in \textsc{Fable}-SE (top) and \textsc{Fable} (bottom), for different FoF linking lengths and gas overdensity thresholds. Solid purple lines show an overdensity threshold of 200, while black dashed, dotted and dash-dotted lines show a threshold of 500. All selections require at least 16 gas cells. The distributions combine detections from snapshots 115, 120, 125 and 130 and are normalised separately for each selection.
    }
    \label{fig:finder_surface_density}
\end{figure}

Figure~\ref{fig:finder_surface_density} shows the gas surface-density distributions of stellar candidates for the different linking lengths and overdensity thresholds, adopting our fiducial minimum of 16 gas cells. At fixed linking length, increasing the overdensity threshold from 200 to 500 has little effect on the distributions. In contrast, the linking length has a much stronger impact. Smaller linking lengths identify smaller, denser structures, shifting the distributions towards higher gas surface densities. This behaviour is similar in both simulations.

\begin{table*}
\centering
\caption{
Sensitivity of stellar-candidate properties to the clump-finding parameters in \textsc{Fable}-SE and \textsc{Fable}, adopting a minimum of 16 gas cells. We show the median logarithmic stellar mass, half-mass radius and mean stellar surface density. The full sample includes measured stars and starless proxies ($\epsilon_\star=0.05$), while the stellar-only sample excludes proxies. Masses, radii and surface densities are in units of $\mathrm{M}_\odot$, pc and $\mathrm{M}_\odot\,\mathrm{pc}^{-2}$, respectively. Counts refer to detections across four snapshots, rather than independent tracks.
}
\label{tab:stellar_finder_sensitivity}
\footnotesize
\setlength{\tabcolsep}{12pt}
\begin{tabular}{cc|rrrr|rrrr}
\hline
 & & \multicolumn{4}{c|}{All stellar candidates} & \multicolumn{4}{c}{Measured stars only} \\
OD & $\ell_{\rm FoF}$ & $N$ & $\log M_\star$ & $\log R_\star$ & $\log\Sigma_\star$ & $N_\star$ & $\log M_\star$ & $\log R_\star$ & $\log\Sigma_\star$ \\
 & (kpc) & & & & & & & & \\
\hline
\multicolumn{10}{l}{\textbf{\textsc{Fable}-SE}} \\
200 & 0.25 & 130 & 6.65 & 2.66 & 0.90 & 83 & 6.83 & 2.78 & 0.91 \\
200 & 0.50 & 1279 & 6.56 & 2.92 & 0.35 & 717 & 6.73 & 3.08 & 0.20 \\
200 & 1.00 & 1265 & 6.68 & 3.21 & -0.06 & 815 & 6.82 & 3.38 & -0.29 \\
500 & 0.25 & 130 & 6.65 & 2.66 & 0.90 & 83 & 6.83 & 2.78 & 0.91 \\
500 & 0.50 & 1264 & 6.56 & 2.92 & 0.35 & 710 & 6.73 & 3.08 & 0.20 \\
500 & 1.00 & 1166 & 6.67 & 3.16 & -0.00 & 705 & 6.84 & 3.35 & -0.19 \\
\hline
\multicolumn{10}{l}{\textbf{\textsc{Fable}}} \\
200 & 0.25 & 88 & 6.64 & 2.72 & 0.86 & 62 & 7.15 & 2.82 & 1.10 \\
200 & 0.50 & 761 & 6.60 & 2.97 & 0.30 & 471 & 6.84 & 3.12 & 0.20 \\
200 & 1.00 & 563 & 6.76 & 3.36 & -0.17 & 425 & 7.01 & 3.44 & -0.28 \\
500 & 0.25 & 88 & 6.64 & 2.72 & 0.86 & 62 & 7.15 & 2.82 & 1.10 \\
500 & 0.50 & 755 & 6.60 & 2.97 & 0.30 & 470 & 6.83 & 3.12 & 0.20 \\
500 & 1.00 & 559 & 6.75 & 3.30 & -0.11 & 394 & 7.07 & 3.43 & -0.19 \\
\hline
\end{tabular}
\end{table*}

\begin{figure*}
    \centering
    \includegraphics[width=0.92\textwidth,keepaspectratio]{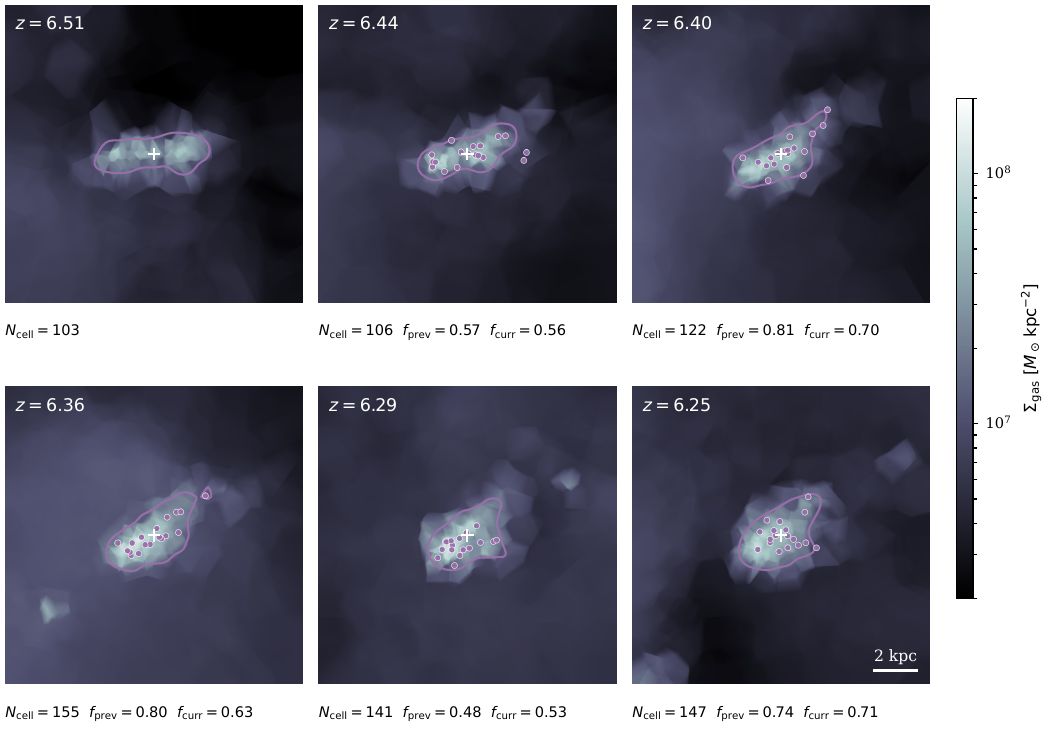}
    \caption{
    Example of clump tracking in \textsc{Fable}-SE across six consecutive snapshots, from $z=6.51$ to $z=6.25$. Each panel shows the gas surface density in a $14\times14$ physical kpc region centred on the tracked clump (ID 1748), with a line-of-sight depth of 12 physical kpc. Purple contours indicate the projected distribution of the clump's constituent mesh elements, while circles highlight a subset with identifiers shared with the previous snapshot. The number of cells and overlap fractions are indicated below each panel.
    }
    \label{fig:tracking_example}
\end{figure*}
We further compare the stellar properties of these candidates in Table~\ref{tab:stellar_finder_sensitivity}. The median stellar masses change relatively little across the tested parameters, generally remaining within a few tenths of a dex. Stellar radii and surface densities are more sensitive to the linking length, with smaller linking lengths producing smaller radii and higher surface densities. The same trends are found when restricting the sample to clumps containing actual stellar particles, excluding our starless proxies.

The absolute number of identified clumps also depends on the adopted parameters. In particular, the 0.25 kpc linking length retains substantially fewer structures, and the enhanced metal-rich population in \textsc{Fable}-SE is not consistently recovered for every selection at this linking length. However, for linking lengths of 0.5--1 kpc, \textsc{Fable}-SE consistently contains more metal-rich clumps than \textsc{Fable} across the tested overdensity thresholds and minimum cell counts. Increasing the minimum cell count primarily removes lower-mass detections. We therefore retain our fiducial choice of 16 gas cells and a 1 kpc linking length, while acknowledging that the absolute clump sizes and densities remain sensitive to the identification method.

\subsection{Clump tracking}
\label{app:tracking}

We also test the robustness of our clump tracking by examining the overlap of AREPO mesh-cell identifiers between consecutive snapshots. Across snapshots 81--130, we identify 4,859 accepted matches in \textsc{Fable} and 7,954 in \textsc{Fable}-SE. The median overlap fractions are 64 and 63 per cent relative to the current clump, and 73 and 68 per cent relative to the previous clump, respectively. These values are substantially higher than our adopted matching thresholds of 5 and 2 per cent.

Increasing the thresholds to 20 and 10 per cent retains more than 99 per cent of the original matches in both simulations. We also independently reconstruct the tracks using thresholds of 5/2, 10/5 and 20/10 per cent. This reproduces 97.4 per cent of the original consecutive links in \textsc{Fable} and 97.9 per cent in \textsc{Fable}-SE. The outward-moving fraction changes only slightly, from 29.1 to 28.2 per cent in \textsc{Fable}-SE and from 11.8 to 10.6 per cent in \textsc{Fable} between the fiducial and strictest diagnostic reconstructions. The inside-to-outside virial-radius population also remains unchanged, with 52 tracks in \textsc{Fable}-SE and none in \textsc{Fable} under all tested thresholds.

We additionally inspect individual tracks visually. Figure~\ref{fig:tracking_example} shows an example in \textsc{Fable}-SE across six consecutive snapshots. The clump remains identifiable, although its morphology and constituent gas cells evolve with time.

These tests show that the main difference in radial evolution between the simulations is insensitive to the adopted matching thresholds. This is further supported by the instantaneous radial-velocity analysis in Section~\ref{sec:radial_migration}, which does not require reconstructed tracks. However, mesh-cell identifiers are not Lagrangian tracers, and we therefore follow the evolution of identified gas overdensities rather than the same gas parcels.

\section{Dependence on the adopted star formation efficiency}
\label{sec:SFproxy}

\begin{figure}
    \centering
    \includegraphics[width=0.48\textwidth,keepaspectratio]{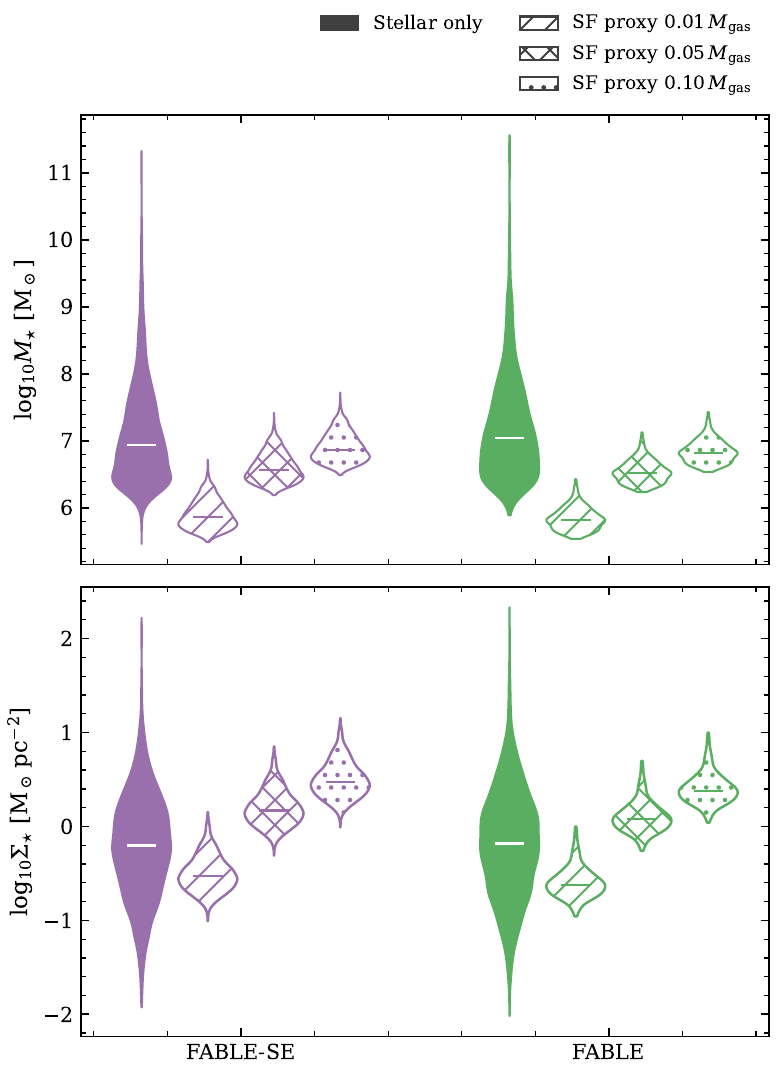}
    \caption{
    Stellar mass (top) and stellar surface density (bottom) distributions in \textsc{Fable}-SE (purple) and \textsc{Fable} (green). Filled violins show clumps containing stellar particles, while hatched violins show the starless proxy sample for assumed star formation efficiencies of 1, 5 and 10 per cent. Horizontal bars indicate the median of each distribution.
    }
    \label{fig:SF_proxies_violin}
\end{figure}

For starless stellar candidates, we estimate the stellar mass assuming a fixed star formation efficiency of $\epsilon_\star=0.05$. To test this assumption, we also consider efficiencies of 1 and 10 per cent. Figure~\ref{fig:SF_proxies_violin} compares the resulting stellar mass and surface density distributions with those of clumps containing actual stellar particles.

As expected, increasing $\epsilon_\star$ shifts the estimated stellar masses and surface densities towards higher values. The fiducial choice of 5 per cent provides an intermediate estimate, with distributions closer to those of the stellar-only population than the lowest-efficiency case. However, this comparison does not independently constrain the star formation efficiency of the unresolved structures.

Importantly, changing $\epsilon_\star$ only affects the inferred stellar properties of the starless candidates. The gas clump identification, stellar-candidate selection, metallicities and radial evolution remain unchanged. Our main conclusions are therefore independent of the assumed efficiency.

\section{The \(T-n_{\rm H}\) plane}
\label{app:tnh_plane}
\begin{figure*}
    \centering
    \includegraphics[width=0.9\textwidth]{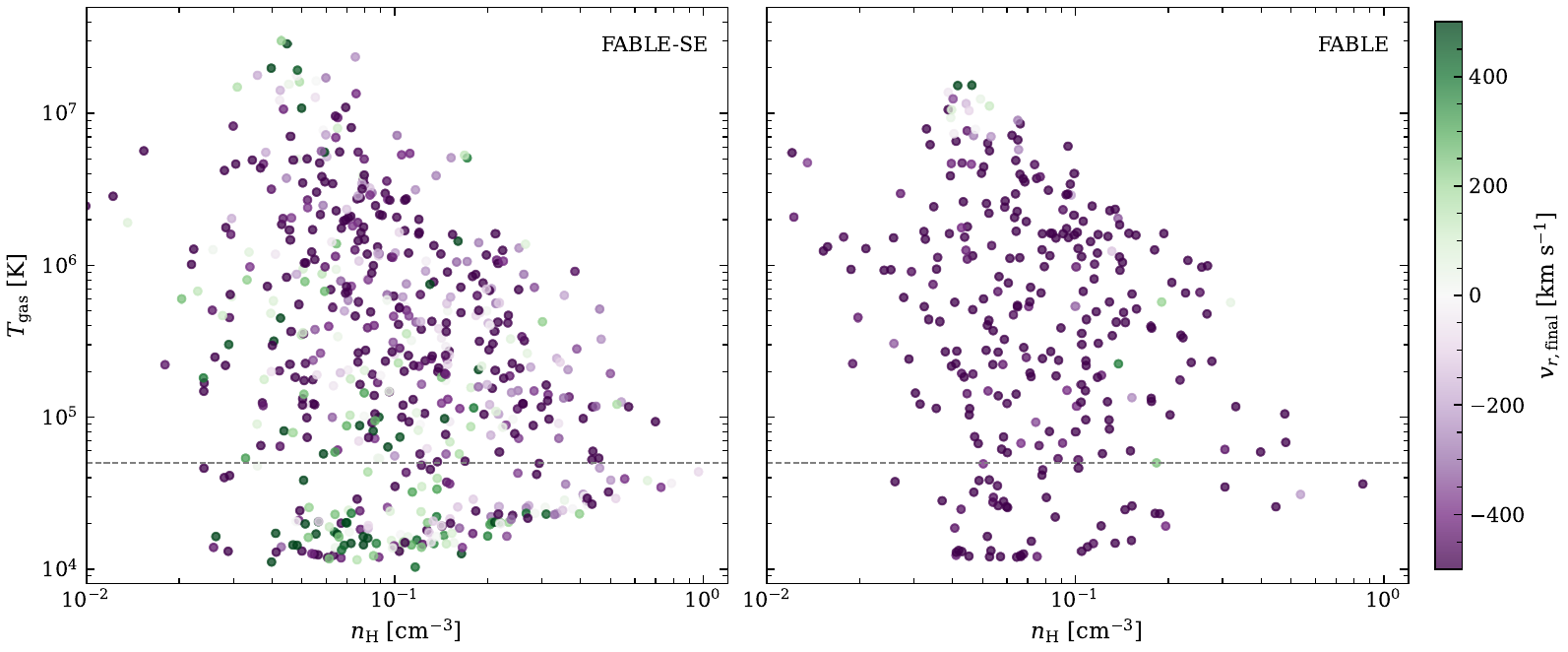}
    \caption{ Gas temperature-density plane for stellar candidates born at $z_{\rm birth}<7$. Points show the median gas temperature, $T_{\rm gas}$, and hydrogen number density, $n_{\rm H}$, measured within each clump, colour-coded by final radial velocity. }
    \label{fig:tnh_plane}
\end{figure*}

The outward-moving \textsc{Fable}-SE stellar candidates have substantially higher cold gas fractions than their counterparts in \textsc{Fable}, as shown in Table~\ref{tab:temperature_properties}. Figure~\ref{fig:tnh_plane} shows the median gas temperature and hydrogen number density of stellar candidates born at $z_{\rm birth}<7$, colour-coded by their final radial velocity. The \textsc{Fable}-SE sample contains more systems with positive final radial velocities than \textsc{Fable}. These objects span a broad range in the $T$--$n_{\rm H}$ plane, including relatively cool and dense gas structures. However, the temperatures of gas on the effective equation of state do not represent resolved physical temperatures, and our cold gas definition includes all such gas regardless of temperature. The figure therefore provides additional information on the gas conditions, rather than a direct measurement of the cold gas fraction used in Table~\ref{tab:temperature_properties}.

%%%%%%%%%%%%%%%%%%%%%%%%%%%%%

% Don't change these lines
\bsp	% typesetting comment
\label{lastpage}
\end{document}